\documentclass{article}
\usepackage[utf8]{inputenc}
\usepackage[margin=1in]{geometry}

\usepackage{microtype}
\usepackage{graphicx}
\usepackage{subfigure}
\usepackage{booktabs} 
\usepackage{natbib}
\setcitestyle{open={(},close={)}}

\usepackage[colorlinks=true,citecolor=blue]{hyperref}

\usepackage{amsmath,amsthm,amsfonts,amssymb,mathdots,array,mathrsfs,bm,bbm,stmaryrd,graphicx,subfigure,xcolor}
\usepackage{algorithm,algorithmic}
\usepackage{breakcites}

\usepackage[T1]{fontenc}
\usepackage{enumerate}
\usepackage{inputenc}

\usepackage{graphicx} % more modern
\usepackage{subfigure}

\usepackage{booktabs,balance}
\usepackage{rotating}
\usepackage{boldline}
\usepackage{makecell}
\usepackage{multirow}
\usepackage{balance}

\usepackage{tikz}

\newtheorem{theorem}{Theorem}[section]

\newtheorem{lemma}[theorem]{Lemma}
\newtheorem{proposition}[theorem]{Proposition}

\newtheorem{remark}{Remark}[section]

\newcommand{\bsmat}{\left[\begin{smallmatrix} }
	\newcommand{\esmat}{\end{smallmatrix}\right] }

\usepackage{environ}
\NewEnviron{smallequation}{%
    \begin{equation}
    \scalebox{0.97}{$\BODY$}
    \end{equation}
    }
    \NewEnviron{smallalign}{%
    \begin{equation}
    \scalebox{0.97}{$\BODY$}
    \end{equation}
    }

\begin{document}

\title{\bf \huge A Cover Time Study of a non-Markovian Algorithm }

\author{\vspace{0.5in}\\
\textbf{Guanhua Fang} \\
School of Management\\
Fudan University, Shanghai, China\\
  \texttt{fanggh@fudan.edu.cn}\\\\
\textbf{Gennady Samorodnitsky} \\
School of Operations Research and Information Engineering\\
Cornell University, New York, USA\\
  \texttt{gs18@cornell.edu}\\\\
  \textbf{Zhiqiang Xu} \\
Department of Machine Learning\\
Mohamed bin Zayed University of Artificial Intelligence, Abu Dhabi, UAE\\
\texttt{Zhiqiang.Xu@mbzuai.ac.ae}\\\\
}

\date{}
\maketitle

\begin{abstract}\vspace{0.2in}
\noindent
Given a traversal algorithm, cover time is the expected number of steps needed to visit all nodes in a given graph.
A smaller cover time means a higher exploration efficiency of traversal algorithm.
Although random walk algorithms have been studied extensively in the existing literature, there has been no cover time result for any non-Markovian method. 
In this work, we stand on a theoretical perspective and show that the negative feedback strategy (a count-based exploration method) is better than the naive random walk search. In particular, the former strategy can locally improve the search efficiency for an arbitrary graph. It also achieves smaller cover times for special but important graphs, including clique graphs, tree graphs, etc.
Moreover, we make connections between our results and reinforcement learning literature to give new insights on why classical UCB and MCTS algorithms are so useful. 
Various numerical results corroborate our theoretical findings.
\end{abstract}

\newpage

\section{Introduction}
\label{sec:intro}

The cover time of a walk/an algorithm on a graph is the expectation of the number of steps required to visit every nodes/vertices. Formally, given a finite graph, we say there is a (directed) edge between node $i$ and node $j$ if we can take some action such that the agent could transit from state $i$ to state $j$. 
For time $n = 0, 1, \ldots$, we use $X_n$ to denote a sequence of nodes covered by the traversal algorithm. 
We define 
\begin{eqnarray}\label{eq:Tc}
T_C &:=& \text{the smallest}~ n ~ \text{such that}~ X_0, X_1, \ldots X_n~ ~\text{visit all nodes of graph,} 
\end{eqnarray}
whose expectation $\mathbb E[T_C]$ is called the \textit{cover time} \citep{broder1989bounds, kahn1989cover}.

It is of particular interest to study the cover time since it quantifies how fast/effectively a walk/an algorithm can traverse the whole graph.
One of the most common but important walk is known as (simple) random walk \cite{pearson1905problem, abdullah2012cover, spitzer2013principles}, which is a sequence of movements from one node to
another where at each step an edge is chosen uniformly at random from the set of
edges incident on the current node, and then transitioned to the next node. 
Cover time on random walk has been studied extensively in past several decades \cite{aldous1991random, lovasz1993note, grassberger2017fast, dembo2021limit}. Other extended types of random walk, including lazy random walk \cite{avin2008explore} and weighted random walk \cite{abdullah2012cover}, have also been considered in the literature.
Unfortunately, all such theoretical results on cover time pertain to memory-less random walk. There has been no result on the cover time of any non-Markovian traversal algorithm.
In other words, it is relatively hard to analyze the covering property of history-dependent random walk algorithms when the Markovian property fails to hold.

In this paper, we try to bridge the aforementioned gap. Specifically, we consider a simple but important non-Markovian traversal algorithm, which we call the \textit{negative feedback} strategy. 
To be more mathematically clear, the negative feedback algorithm is a count-based method. If $X_n = i$ at time $n$, the next state is uniformly randomly selected from a subset $Smin_i^{(n)}$ of $i$'s neighbours, where $Smin_i^{(n)}$ contains those nodes which are \textit{least} visited from state $i$ up to time $n$. 
Such procedure is called the \textbf{"favor least"} mechanism. Heuristically, it tends to move to un/less-visited nodes and hence can improve the cover time.
Undoubtedly, it is history-dependent, since it requires counting transitions between each pair of neighbour nodes.

Why should we consider the negative feedback algorithm? Reasons are of three-fold. First, it is one of the simplest non-Markovian random walk algorithms. There is less hope of making any theoretical claims on the cover time of a very complex traversal algorithm. 
Second, it only requires to store a count table where each entry represents the number of movements from one node to its neighbor. It can be updated very efficiently and fast.
Third, it has strong connections with algorithms in reinforcement learning (RL) field. To be more concrete, given a discrete-state environment, we can treat each state as a node. 
The agent can take a certain policy to explore the whole environment. 
The negative feedback strategy is often treated as an exploration tool \cite{mcfarlane2018survey, hazan2019provably} in an unknown Markov Decision Process (MDP) setting.

Our main results in this work are summarized here. 
\begin{itemize}
    \item[I]. 
    We first show a local improvement result for a general graph. 
To be specific, we consider a local version of the negative feedback algorithm where the "favor least" mechanism is only applied to the starting state $X_0$.
For an \textbf{arbitrary} graph, we have shown that $\mathbb E_{\pi_{loc}}[N_j|X_0] \leq \mathbb E_{\pi_{rw}}[N_j|X_0]$ for any other node $j \neq X_0$,
where $N_j$ is defined to be the number of excursions outside starting node $X_0$ before state $j$ is visited for the first time,
$\pi_{loc}$ and $\pi_{rw}$ stand for the local negative feedback algorithm and random walk policy, respectively.
This local improvement result implies that the negative feedback mechanism improves the exploration efficiency, at least locally, in the sense that the agent has the stronger tendency to visit other nodes instead of returning to the starting node.
    \item[II]. 
    We then make a step forward and show cover time improvement under several special graph structures.
We are able to show that 
$\mathbb E_{\pi_{neg}}[T_C] < \mathbb E_{\pi_{rw}} [T_C]$ ($\pi_{neg}$ represents the negative feedback algorithm) under Star, Path, Clique and Tree graphs.
In particular, in the case of a \textbf{balanced $b$-ary tree}, we establish that 
$\mathbb E_{\pi_{neg}}[T_C] \leq 4 H \frac{b+1}{b-1} b^H$, where $b$ is the number of children of each non-leaf node and $H$ is the depth of tree.
By contrast, it was shown \citep{aldous1991random} that $\mathbb E_{\pi_{rw}} [T_C] \approx 2 H^2 b^{H+1} \frac{\log b}{b-1}$.
Therefore, the negative feedback algorithm improves the cover time by order of $H \log b$.
In other words, in tree-like RL games, the naive random walk search becomes less efficient compared with negative feedback strategy as action and state spaces become more complex.
\end{itemize}

The rest of the paper is organized as follows.
In Section \ref{sec:2}, we introduce the cover time formulation and provide an illustrative example to show why the negative feedback algorithm
can improve over the random walk policy.
In Sections \ref{sec:3} and \ref{sec:4}, we establish the local and cover time improvement results, respectively.
In Section \ref{sec:5}, we make connections to the maximum-entropy exploration, UCB, and Monte Carlo tree search methods and make attempts on non-discrete cases.
A concluding remark is given in Section \ref{sec:conclusion}.
Numerical experiments, additional discussions and technical proofs are provided in the appendices.

\vspace{-1mm}

\section{Cover Time Formulation}\label{sec:2}

\vspace{-1mm}

Let us imagine that an agent walks on a finite and connected graph. 
If an agent could take some action to transit from node $i$ to node $j$, then we say there is a (directed) edge from node $i$ to $j$.
To help reader understand the terminology clearer, we provide the following examples.
In a two-dimensional Grid Word environment, a node can be viewed as the position of the agent.
Since the agent can choose to move up, down, right or left, two nodes have an edge between them if and only if these two positions are adjacent to each other.
In a Go game, two players take turns to put stones on the board. A node here represents a 19 $\times$ 19 board with white and black stones on it.
There is a directed edge from node $i$ to another node $j$, only if node $j$ can be reached from node $i$ after a player takes a single action.

Given a starting time $n = 0$ and an initial node $X_0$, we let 
$X_n, \, n=0,1,\ldots$ describe the sequence of states/nodes governing by some exploration strategy and denote 
\begin{eqnarray} \label{e:C}
T_C &=&  \text{the first time $n$, $X_0, X_1, \ldots, X_n$} ~ \text{visit all nodes
	of the graph.}
\end{eqnarray}
The quantity $\mathbb E[T_C]$ (expectation of $T_C$) is called the cover time \citep{broder1989bounds, kahn1989cover}. 
In this paper, we mainly focus on two  exploration strategies, random walk algorithm \citep{aldous1991random, dembo2021limit} and negative feedback algorithm, whose formal mathematical formulations are described as below. 

\noindent {\bf Random walk algorithm} \ This is a Markovian mechanism; if
$X_n=i$ at some time $n=0,1,2,\ldots$ for some node $i$, the next state
is chosen uniformly at random among the neighbours of $i$ in the
graph. 
\begin{eqnarray} \label{e:RW.tp}
&&\mathbb P(X_{n+1}=j|X_0=i_0,\ldots, X_{n-1}=i_{n-1},X_n=i) \nonumber \\
&=&
\left\{ \begin{array}{ll}
1/d_i & \text{if $(i,j)$ is an edge} \\
0 & \text{if $(i,j)$ is not an edge},
\end{array}
\right.
\end{eqnarray}
where $d_i$ is the degree of the node $i$.

\noindent
{\bf Negative feedback algorithm} \ For every node $i$ of the
graph and every neighbour $j$ of $i$ let $N_{ij}^{(n)}$ be the number
of times the agent moved from node $i$ to node $j$ prior to time
$n$ (so that $N_{ij}^{(0)}=0$ for all nodes $i,j$) and denote
\begin{align} \label{e:Nmin}
&Nmin_{i}^{(n)}=\min_{j:\, (i,j)\ \text{an edge}} N_{ij}^{(n)}, \\  
\notag   &Smin_{i}^{(n)}= \bigl\{  j:\, (i,j) \ \text{is an edge and} \ 
N_{ij}^{(n)} =  Nmin_{i}^{(n)}\bigr\},  \\
& K min_{i}^{(n)} = \
\text{cardinality}(Smin_{i}^{(n)}). \notag
\end{align}
Then $X_n=i$ at some time $n=0,1,2,\ldots$ for some vertex $i$, the
next state
is chosen uniformly at random among the neighbours of $i$ in the
graph with the smallest prior selection. That is,
\begin{eqnarray} \label{e:NF.tp}
&& \mathbb P(X_{n+1}=j|X_0=i_0,\ldots, X_{n-1}=i_{n-1},X_n=i)  \nonumber \\
&=&
\left\{ \begin{array}{ll}
1/K min_{i}^{(n)} & \text{if} \ j\in Smin_{i}^{(n)} \\
0 & \text{otherwise}. 
\end{array}
\right.
\end{eqnarray}
In other word, a neighbour node $j$ will be never chosen only if it becomes the least visited one from current node $i$. 
It is not hard to see that negative feedback algorithm requires storing the counts of transitions between pair of nodes with edge among them.
Therefore the algorithm is non-Markovian and does not have nice properties (e.g. regeneration property) as random walk algorithm does.
It makes theoretical analysis extremely hard in studying general graph.

\begin{remark}
There are quite a few existing works on the cover time of random walk algorithm (see \cite{kahn1989cover, feige1995tight, abdullah2012cover} and references therein) and its variants (e.g. lazy random walk \citep{avin2008explore}, random walk
with heterogeneous step lengths \citep{guinard2020tight}).
However, to our knowledge, there is no literature considering the cover time problem of any count-based algorithm.
\end{remark}

Let us first consider the following specific toy example, which shows the advantage of negative feedback algorithm over random walk algorithm.   

\begin{figure}[ht!]
	\centering
\includegraphics[width=0.45\textwidth]{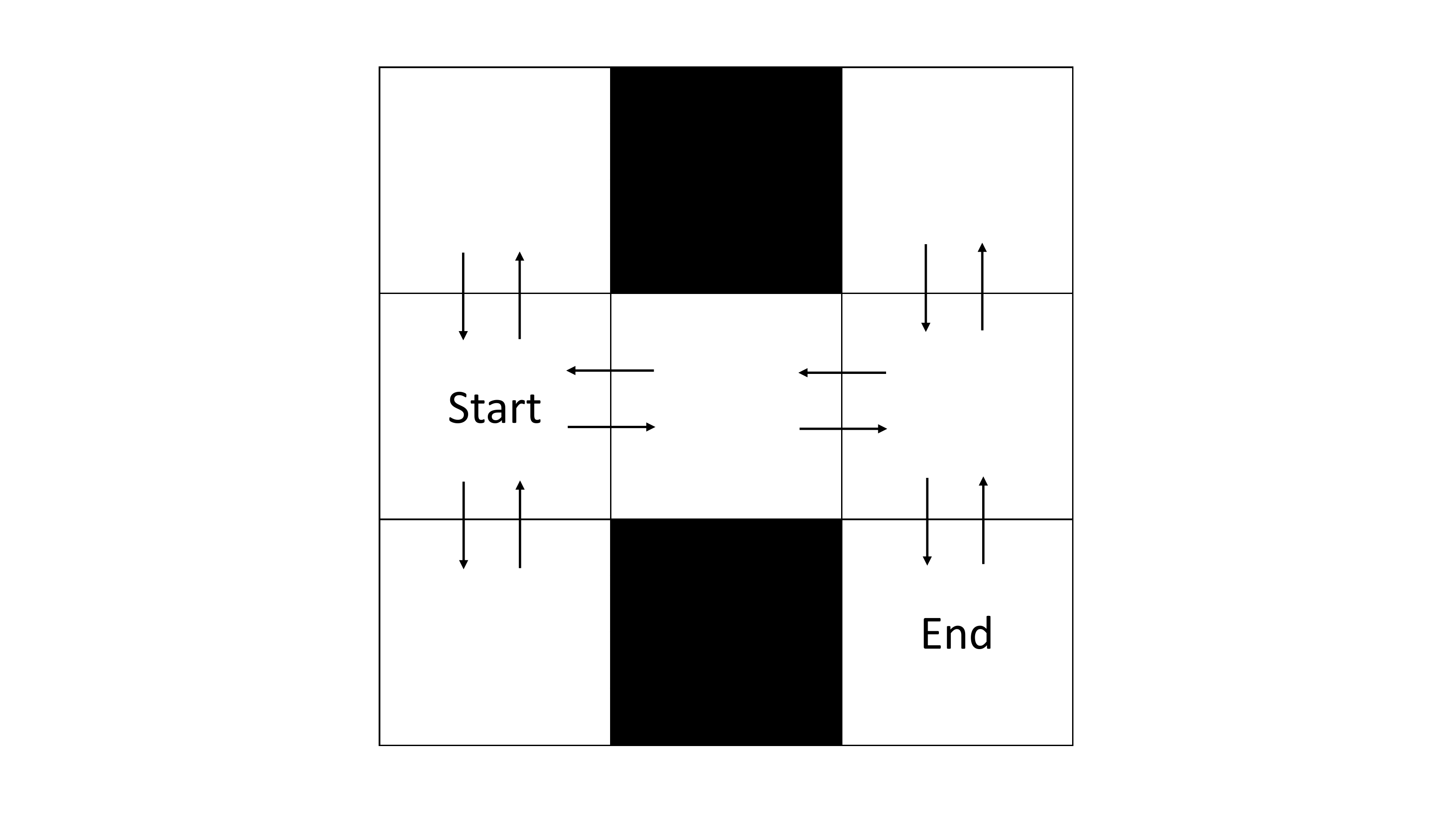}
	\caption{A 3 by 3 grid world. The black grid is inaccessible. The arrow represents the action can be taken in each grid. }\label{fig:badeg}
\end{figure}

\noindent \textbf{A toy grid world}.~ It is a three by three two-dimensional maze as shown in Figure \ref{fig:badeg}. Black grids are obstacles which are not accessible.
At starting time $n = 0$, the agent is placed at "Start" grid. 
The "End" grid is the target node. 
The positive reward will not be given until the agent arrives at "End" grid.
One would wonder, under which policy between random walk and negative feedback algorithms, the agent can take fewer steps in this simple task?

We define 
$T_{task}^{\pi}$ to be the first time of arriving "End" under a policy $\pi$. 
Let $\pi_{rw}$ be the random walk policy
and $\pi_{neg}$ be the negative feedback algorithm.
Is $\mathbb E[T_{task}^{\pi_{neg}}] < \mathbb E[T_{task}^{\pi_{rw}}]$? 
The answer is affirmative as indicated by the following proposition.

\begin{proposition}\label{prop:1}
    In the toy grid world described above, we have 
    $\mathbb E[T_{task}^{\pi_{neg}}] < \mathbb E[T_{task}^{\pi_{rw}}] \equiv 23$.
\end{proposition}

Moreover, we consider the \textit{temporally-persistent/extended} policy \citep{dabney2020temporally}, where one can randomly choose an action and perform it for consecutive times.
To be specific, the agent first choose the direction (up, down, right or left) uniformly randomly and then choose the repetition time $z \sim p(z)$ $(z = 1, 2, \ldots)$. 
We denote such policy as $\pi_{per}(p)$.
When $p(z = 1) =1$, the policy reduces to the random walk strategy.
In this toy grid word, it can be shown that the temporally-persistent strategy is even worse than random walk method. That is,
$\mathbb E[T_{task}^{\pi_{per}(p)}] \geq \mathbb E[T_{task}^{\pi_{rw}}]$.

\begin{proposition}\label{prop:2}
    In the same toy grid world as in Proposition \ref{prop:1}, we have 
    $\mathbb E[T_{task}^{\pi_{rw}}] \leq \mathbb E[T_{task}^{\pi_{per}(p)}]$ for any distribution $p$.
\end{proposition}

\section{Local Improvement}\label{sec:3}
 
Our goal is to provide a theoretical explanation why one can expect
that, at least in some cases, the negative feedback algorithm has a
smaller cover time than the random walk. 
Since a direct analytical
analysis of the cover time on an arbitrary graph is extremely difficult, we will instead look at a related
quantity in this section.

For a node $j$ of the graph, we denote the first hitting time of $j$ by
$$
T_j=\inf\{ n \geq 1:\, X_n=j\}.
$$
For an arbitrary node index $i$,
$\mathbb E[T_j|X_0=i]$ represents the expected time to reach vertex $j$ starting in
vertex $i$. It is intuitively clear that the quantities $\bigl(
\mathbb E[T_j|X_0=i], \, i,j ~\text{are two nodes} \bigr)$ are strongly related to the cover time. 
More precisely, we define
\[
\mu^+=\max_{i,j} \mathbb E[T_j|X_0=i], ~~ \mu^-=\min_{i,j} \mathbb E[T_j|X_0=i].
\]
Then for any starting node $i$ of the graph, it holds
\begin{equation} \label{e:cover.bounds}
\mu^-H_{m-1}\leq \mathbb E[T_C] \leq \mu^+H_{m-1},
\end{equation}
where $m$ is the total number of nodes in the graph and
$$
H_k := 1+1/2+\cdots +1/k
$$
is the $k$th harmonic number; see \cite{matthews1988covering} for more detailed explanations. Therefore, as
a substitute for comparing directly the cover time under the
negative feedback algorithm and the random walk, it is desired to
compare the expected first hitting times under these two algorithms.

For any two arbitrary nodes $i$ and $j$ in graph, a direct
comparison of $\mathbb E[T_j|X_0=i]$ under the two algorithms also seems to be prohibitively difficult. 
We further instead compare another related quantities. 
Let
$V_0=0$ and, for any integer $k\geq 1$, also let
$$
V_k=\inf\bigl\{ n>V_{k-1}:\, X_n=i\bigr\},
$$
the time of the $k$th visit to state $i$.
%assuming
%$V_{k-1}<\infty$.
If $V_{k-1}=\infty$, then we set $V_k=\infty$ as
well. We can think of the time interval $\{V_{k-1}+1,\ldots, V_k\}$ as the
$k$th excursion outside of the vertex $i$. Let
$$
N_j=\inf\bigl\{ k\geq 1:\, T_j\leq V_k\bigr\}
$$
be the number of the excursion outside of $i$ during which vertex $j$
is visited for the first time. It is clear that $\mathbb E[T_j|X_0=i]$ is
also closely related to $\mathbb E[N_j|X_0=i]$.
The intuition is that smaller $N_j$ indicates that agent spends less time to discover node $j$.

In the rest of section, we will aim to compare the
latter quantity between two algorithms. 
To make the comparison
possible and circumvent the non-Markovian issue, we consider the \textbf{local version} of the negative feedback algorithm.

 \noindent
\textit{
 ({\color{purple} Local negative feedback algorithm})
The mechanism \eqref{e:NF.tp} is used {\bf only when the agent is in the starting state $i$}. In every
other states, the random walk dynamics is used! 
}

By such modification, we are able to show that 
values of $\mathbb E[N_j|X_0=i]$ under the local negative feedback algorithm is no larger than that of random walk exploration strategy.

\begin{remark}\label{rmk:tech}
The technical reason of considering local negative feedback algorithm is explained here. 
A nice property of naive random walk strategy is known as \emph{regeneration} property, where everything is reset once the agent returns back to the original state. It makes computation of mathematical recursive formula possible. 
Unfortunately, negative feedback algorithm relies on the past information and does not has such regeneration property. 
Instead, the local version of negative feedback can mitigate this issue. Despite of being non-Markovian, the regeneration can still happen at the time $n$ when $Kmin_i^{(n)} = d_i$.
\end{remark}

\begin{theorem}\label{thm:local}
For any given starting node $i$, it holds 
\begin{eqnarray}\label{eq:local}
\mathbb E_{\pi_{loc}}[N_j|X_0 = i] \leq 
\mathbb E_{\pi_{rw}}[N_j|X_0 = i]
\end{eqnarray}
for any node $j \neq i$ in the graph.
\end{theorem}

Theorem \ref{thm:local} says that, on average/expectation, the local negative feedback algorithm takes less number of excursions (outside the starting state) to visit any other states .
Therefore, a local modification at state $i$ indeed improves the exploration efficiency, i.e., stronger tendency of visiting other states rather than returning back to the initial state.

\section{Cover Time Improvement}\label{sec:4}

As described in the previous sections that direct analytical analysis of cover time is not an easy job for the negative feedback strategy whose non-Markovian mechanism makes computation prohibitively hard.
However, fortunately, we are able to show that the negative-feedback strategy is strictly better than random walk strategy, $\mathbb E_{\pi_{neg}}[T_C] < \mathbb E_{\pi_{rw}}[T_C]$, when the graph admits some special structures.

To start with, we first provide a general property of the negative feedback algorithm, i.e. the worst case of $T_C$ is always bounded. 

\begin{theorem}\label{thm:general}
    For any connected graph, there exists a positive integer $G$ such that $T_c \leq G$. Here $G$ may depend on the number of nodes in the graph, the maximum degree of a single node and the length of the longest path.
\end{theorem}

By Theorem \ref{thm:general}, we know that the negative feedback policy can traverse all the nodes in finite time. By contrast, the random walk policy does not have such property. Mathematically, for any graph with $d_{max} \geq 2$ (where $d_{max}$ is the largest node degree in the graph), it holds 
\begin{eqnarray}
    \mathbb P_{\pi_{rw}} (T_C > N) > 0 ~~~ \text{for any integer}~ N \in \mathbb N. 
\end{eqnarray}

Another immediate conclusion from the above theorem is that $\mathbb E_{\pi_{neg}}[T_C] \leq G$. 
But for most graphs, the constant $G$ obtained in Theorem \ref{thm:general} is too loose. In the proof, we can see that $G$ grows exponentially as the longest path length grows. We may get rid of such exponential relationship after taking into account specific graph structures. 
In the rest of this section, we make efforts to establish tighter bounds for special graphs including "Star", "Path", "Circle", "Clique" and "Tree", which are the most common graphs studied in the literature of cover time analysis for random walk policy.

\begin{figure}[ht!]
	\centering
		\includegraphics[width=0.48\linewidth]{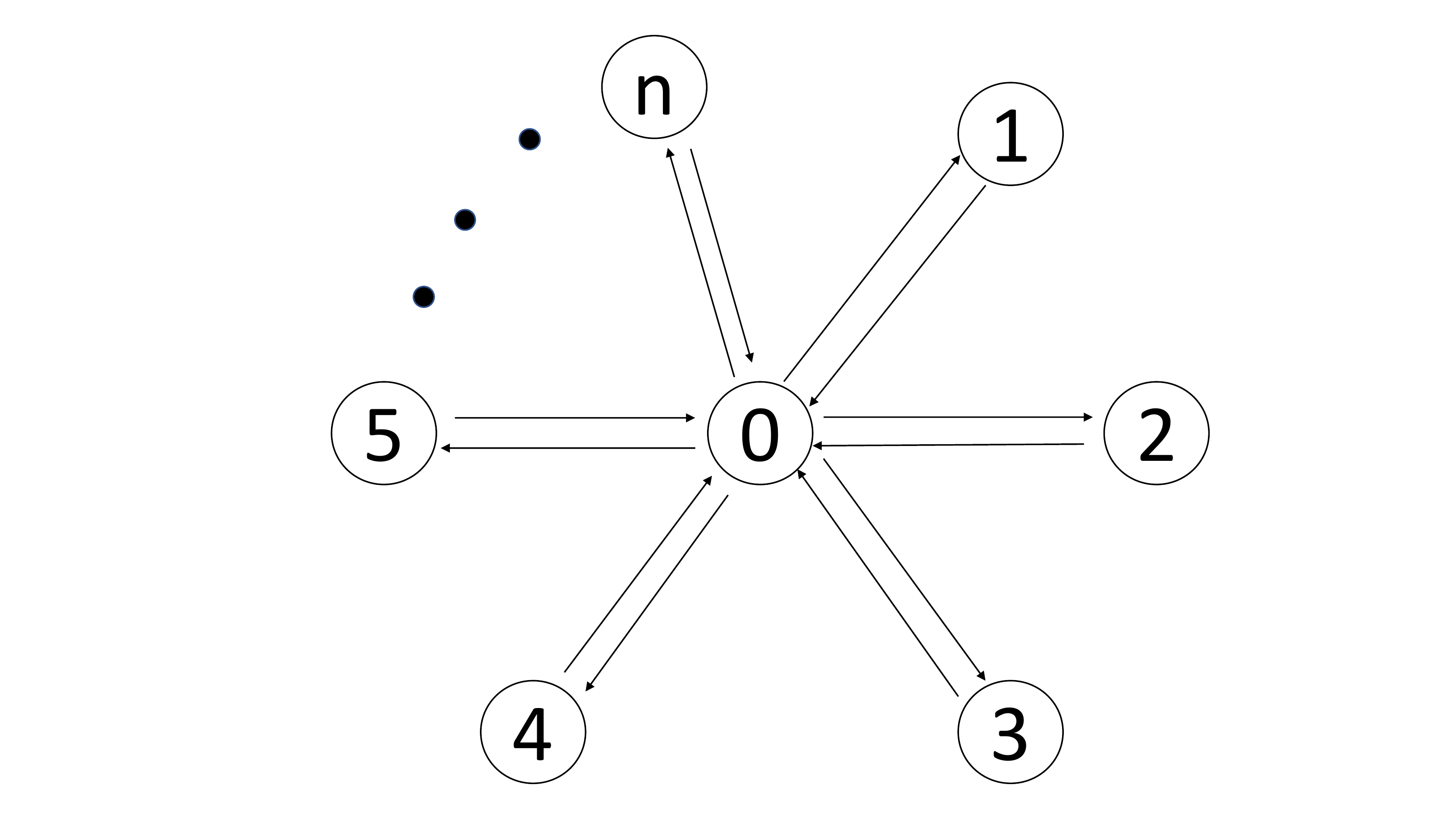}
		\includegraphics[width=0.48\linewidth]{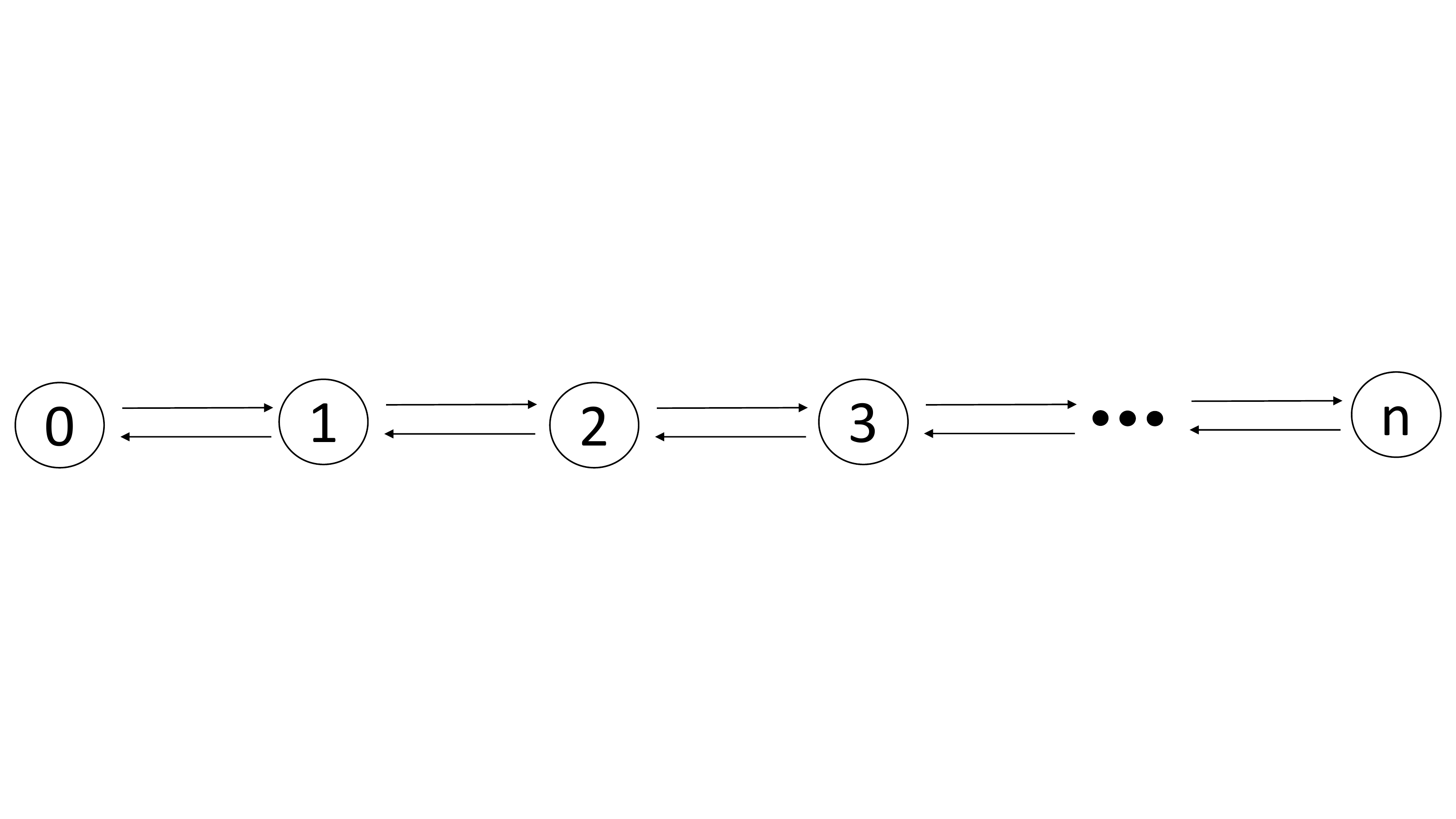}
        \hspace{1mm}
        \includegraphics[width=0.48\linewidth]{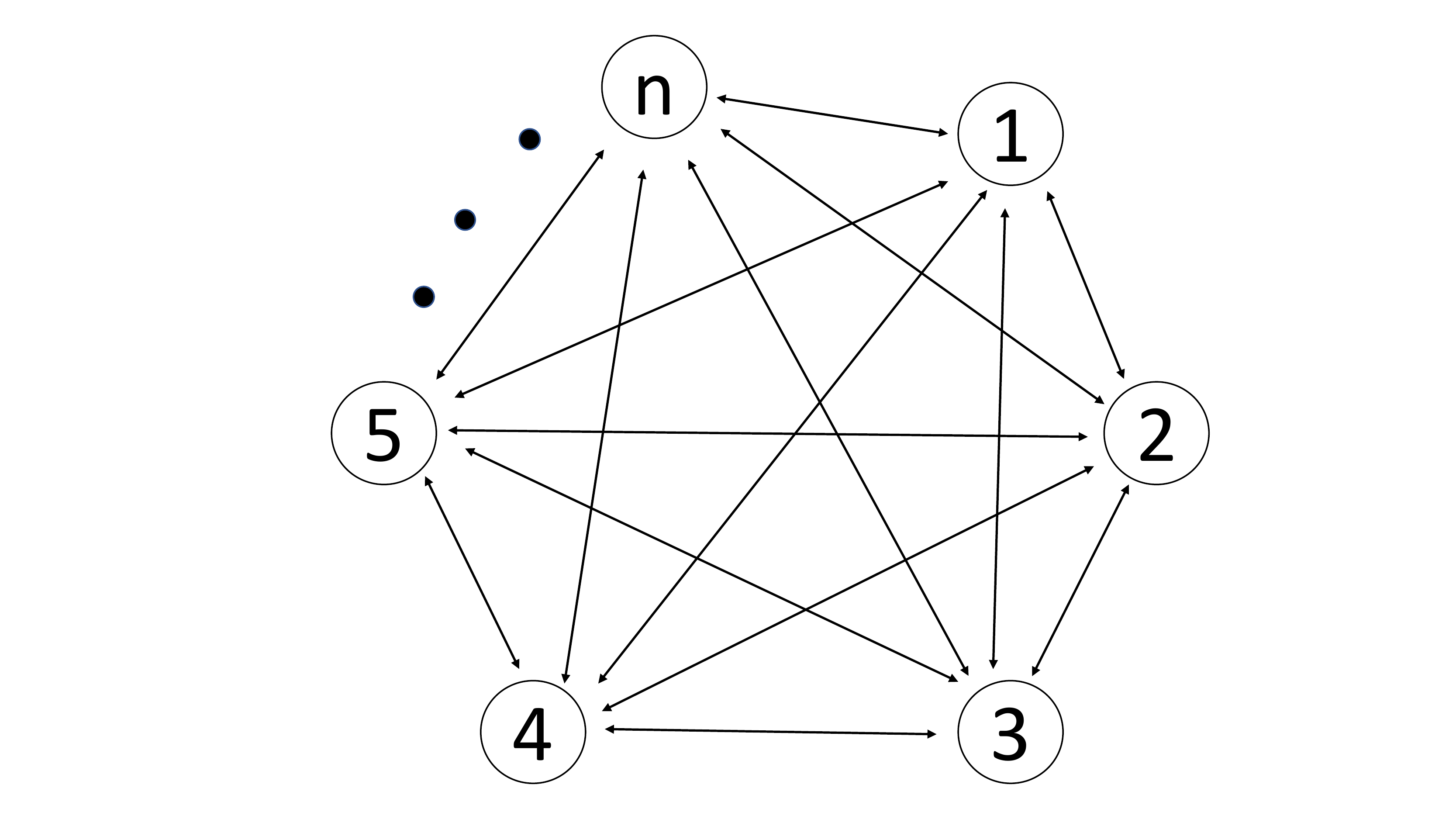}
        
        \includegraphics[width=0.48\linewidth]{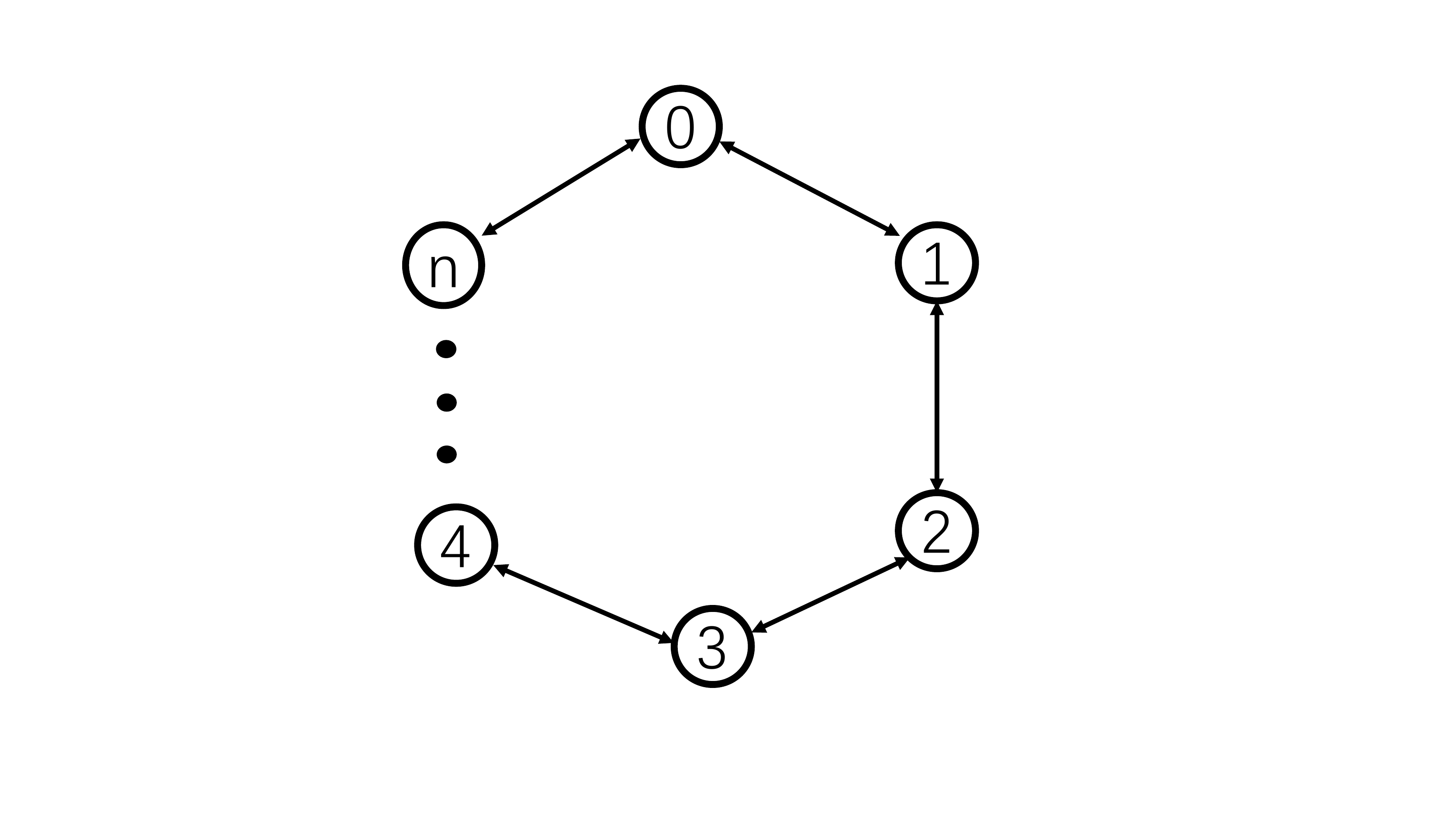}
        \includegraphics[width=0.48\linewidth]{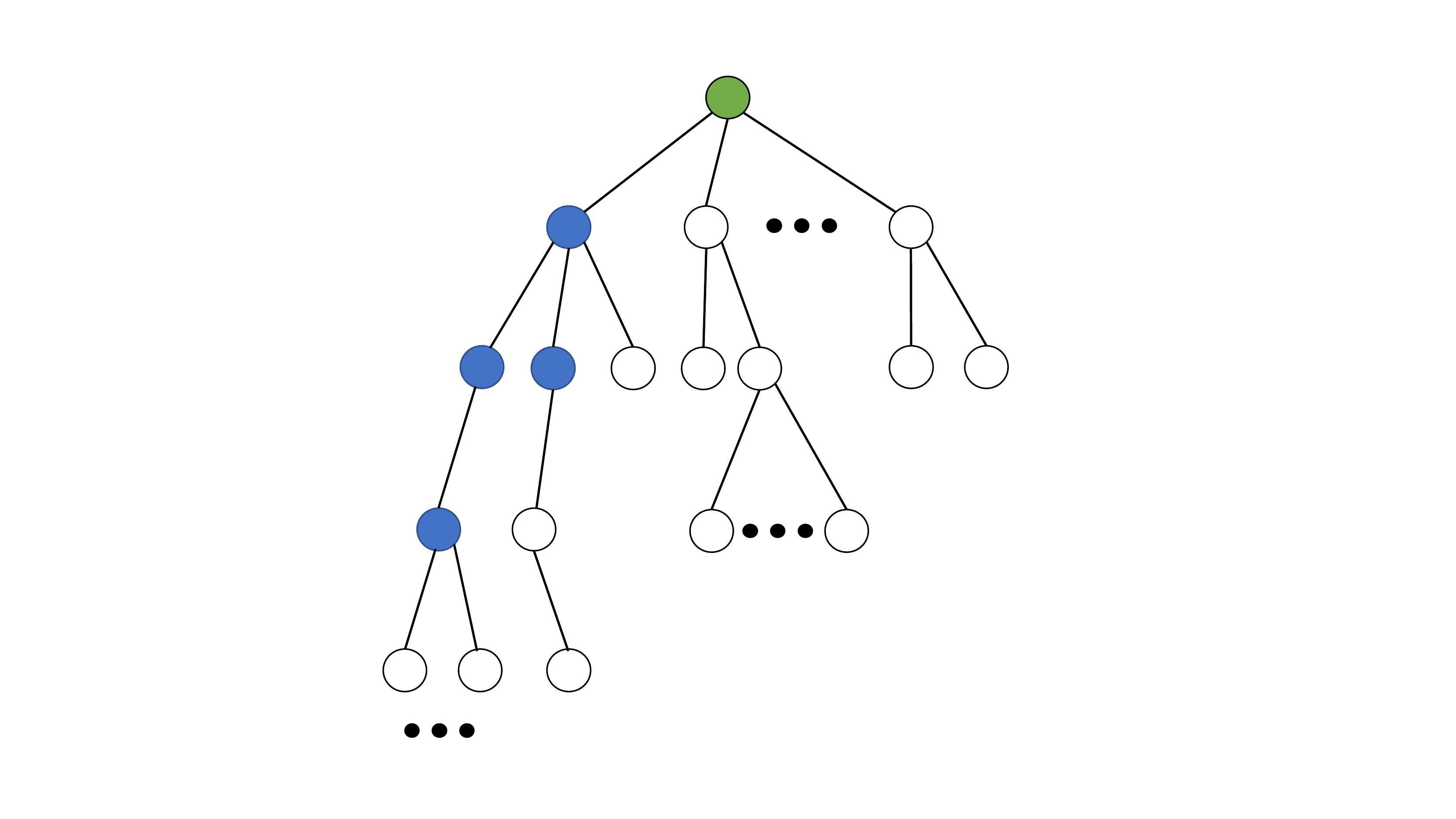}

	\caption{Upper left: Star graph with initial state at 0. Upper right: Path graph with initial state at 0. 
 Center: Clique graph with an arbitrary initial state.
 Bottom left: Circle graph with an initial state at 0. 
 Bottom right: a tree graph with root node as initial state.}\label{fig:star}
\end{figure}

\medskip

\textbf{Star Graph}.~ 
There is an central node (state) 0 and it connects to $n$ leaf nodes. 
The starting position is node 0. 
See Figure \ref{fig:star} for graphical illustration. 
It is easy to see that the degree of node 0 is $n$ and degree of each leaf node is 1.

\begin{theorem}\label{thm:star}
	In the star graph with $n \geq 2$, it holds that $\mathbb E_{\pi_{neg}}[T_C] = 2n - 1$ and $\mathbb E_{\pi_{rw}}[T_C] = 2n (\sum_{i=1}^n \frac{1}{i}) - 1$.
	Hence, cover time of negative feedback policy is strictly smaller than that of random walk policy.
\end{theorem}

\textbf{Path}.~ All $(n+1)$ nodes are aligned in a line. Node $i ~ (\neq 0~ \text{or}~ n)$ is connected to $i-1$ and $i+1$.  The initial state is state 0. In this graph, nodes $0$ and $n$ have degree 1. All other nodes have degree 2.

\begin{theorem}\label{thm:path}
	In the path graph with $n \geq 2$, it holds that $\mathbb E_{\pi_{neg}}[T_C] < n^2 \equiv \mathbb E_{\pi_{rw}}[T_C]$. 
Hence cover time of negative feedback policy is strictly smaller than that of random walk policy.
\end{theorem}

\textbf{Circle}.~ All $(n+1)$ nodes are aligned in a circle. Node $i$ is connected to node $(i-1) \% (n+1)$ and node $(i+1) \% (n+1)$, where $\%$ stands for modulo. The initial state is state 0. In this graph, all nodes have degree 2.

\begin{theorem}\label{thm:circle}
	In the path graph with $n \geq 2$, it holds that $\mathbb E_{\pi_{neg}}[T_C] < \frac{1}{2}(n+1)n \equiv \mathbb E_{\pi_{rw}}[T_C]$. 
Hence cover time of negative feedback policy is strictly smaller than that of random walk policy.
\end{theorem}

\textbf{Clique Graph}.~ It is a graph of $n$ nodes such that any pair of nodes has an edge between them. It is not hard to see that every node has degree $n$.

\begin{theorem}\label{thm:complete}
	In the clique graph with $n \geq 3$, the strict inequality $\mathbb E_{\pi_{neg}}[T_C] < \mathbb E_{\pi_{rw}}[T_C]$ holds.
\end{theorem}

\textbf{Tree Graph}.~ It is a graph of nodes with no circle. The tree has depth $H$ and each non-leaf node has at most $b$ child nodes.

\begin{remark}
In fact, every RL environment can be reformulated as a tree graph if we treat every state-action trajectory as a single node. This idea is usually adopted in counter-factual regret minimization (CFR, \cite{zinkevich2007regret}).
\end{remark}

\textbf{Balanced $b$-ary Tree Graph}.~ It is a tree graph with depth $H$ and each non-leaf node has exactly $b$ children.

It can be seen that the balanced $b$-ary tree is a special case of a general tree. 
In the literature, the following result of cover time under random walk policy has been established in 1990s \citep{aldous1991random}.
\begin{proposition}[\cite{aldous1991random}]\label{prop:rw:tree}
	For a balanced $b$-ary tree of depth $H$, the cover time of random walk algorithm is asymptotically 
	$$ 2H^2 b^{H+1} (\log b)/(b - 1)$$
	as $H \rightarrow \infty$.
\end{proposition}

We first show there exists an upper bound of visiting each node under negative feedback algorithm.
Therefore, it will not waste too much time on visited nodes before all node have been visited at least once.

\begin{theorem}\label{thm:tree}
	In the tree graph, under negative feedback algorithm, each node is visited at most $2 (b+1) H$ times before all nodes have been visited at least once.
\end{theorem}

By counting the total number of nodes, a direct application of Theorem \ref{thm:tree} will lead to that,
in the balanced $b$-ary tree graph with depth $H$, it holds 
$\mathbb E_{\pi_{neg}}[T_C] \leq 2 (b+1) H \frac{b^{H+1} - 1}{b-1}$.

A more refined analysis will give us the following result.
\begin{theorem}\label{thm:tree:all}
    In the balanced $b$-ary tree graph with depth $H$, under negative feedback algorithm, it holds 
	$\mathbb E_{\pi_{neg}}[T_C] \leq 4 H \frac{b+1}{b-1} b^H$.
\end{theorem}

Compared with Proposition \ref{prop:rw:tree}, the negative feedback algorithm is asymptotically $H$ times faster than the random walk algorithm for any fixed $b$.
It is also asymptotically $\log b$ faster for any fixed $H$.
Therefore, the negative feedback algorithm improves the search efficiency in terms of both tree width and tree depth. 
Moreover, there are $\frac{b^{H+1}-1}{b-1}$ nodes in a $b$-ary tree, which indicates that the cover time is no smaller than order of $b^H$. In other words, the negative feedback algorithm visits each node times on average of order $H$, while the random walk algorithm visits each node on average $H^2 \log b$. This is a substantial improvement.
In practice, for many board games which can have large action spaces and form very deep trees, the random walk exploration is a less efficient strategy according to our theoretical explanation from the cover time perspective.

Additionally, $4 H b^H \frac{b +1}{b-1}$ is actually also the worst case bound of $T_C$ (in addition to the bound on expectation $\mathbb E [T_C]$) under the proposed algorithm. By contrast, $2H^2 b^{H+1} (\log b)/(b - 1)$ is only the upper bound of expectation, $\mathbb E[T_C]$.
In other words, with non-vanishing probability (in some extreme cases), $T_C$ can be exponentially large under random walk policy as number of nodes grows.

\section{Connections and Discussions}\label{sec:5}

In previous sections, we have established properties of negative feedback algorithm on the finite-node graph and show that it indeed improves cover times of several important graphs. In below, we make connections to reinforcement learning field and provide practical implications why negative feedback strategy is so interesting and important to many popular RL algorithms.
Before going to detailed discussions, we would like to make the clarification that we are not trying to propose any new RL algorithm in our paper. Negative feedback algorithm considered here is just a counterpart/non-Markovian extension of random walk. It does not rely on any Markov decision process setting or any Bellman equation-related assumptions \cite{puterman1990markov}.
The following discussions are mainly heuristic without mathematical justifications and help readers to realize the importance of the negative feedback strategy.

\subsection{Connection with $\epsilon$-Greedy Methods}

The negative-feedback exploration strategy can be easily incorporated into any existing reinforcement learning algorithm. 
For example, in $\epsilon$-greedy type of methods \citep{sutton1995generalization, wunder2010classes}, we can replace random action selection by using negative-feedback strategy.

With $1 - \epsilon$ probability, agent adopts a learned policy for exploiting the current possible maximum reward.
In the literature, such learner can be either model-based method (UCRL2 \citep{jaksch2010near}, UCRL2B \citep{fruit2020improved}, etc.) or model-free method ($Q$-learning \citep{watkins1992q}, SARSA \citep{sutton2018reinforcement}, etc.). 
With $\epsilon$ probability, exploration policy is used for exploring the entire state space and can help escaping the local optimum. 
As discussed in previous sections, 
the negative feedback algorithm is indeed a better strategy than random work from theoretical perspective.
As a result, negative feedback strategy can theoretically improve the efficiency of any $\epsilon$-greedy-type RL algorithms in tree-like environment.

	% Random walk is the most standard and classical search policy. Its implementation is easy and straightforward. But sometimes it could be inefficient since it spend too multiple times in the exploraing the same state. 
	
	% The \textit{temporally-persistent} (\textit{temporally extended}) policy is proposed by \cite{dabney2020temporally}.
	% It incorporates the temporal persistence. That is, it randomly takes an action $a$ and performs it for $k \in \mathbb N^{+}$ consecutive times. 
	% It can enhance the efficiency in exploration along a single direction.
	% The distribution of $k$ could be a discrete uniform distribution, zeta distribution, etc. 

\subsection{Connection with RL Exploration Methods}

A fundamental problem in reinforcement learning is how to explore the state space faster, especially when the environment only provides sparse rewards or even no reward before reaching the final state. This question has received a lot of attentions, with approaches such as intrinsic-reward learning \citep{chentanez2004intrinsically,bellemare2016unifying}, curiosity-driven algorithm \citep{pathak2017curiosity, burda2018large}, etc.

Among those, maximum-entropy exploration policy
\citep{hazan2019provably} arouses special interests in recent years. The policy needs to iteratively learn the unknown MDP. 
For un-explored / less-explored node $s$ (see definition of $m$-known state in \cite{hazan2019provably}), they select action,
$$\arg\min_a N(s,a),$$
where $N(s,a)$ is the cumulative count number of choosing $a$ at state $s$ up to the current round. In other words, the negative feedback algorithm considered in our paper serves as an important role in learning unknown transition probabilities. Our theories answer that why popular RL exploration methods prefer to using "favor least" mechanism rather than using simple random walk strategy.

\subsection{Connection with UCB method}

The upper confidence bound (UCB) algorithm \citep{auer2002using, auer2002finite} is probably one of the most famous methods in balancing exploration and exploitation. 
At time $n$ and state $s$, the agent chooses the best action according to the following criteria,
\begin{eqnarray}\label{eq:ucb}
    \arg\max_a \big\{ \hat r(s,a) + c \sqrt{\frac{\log n}{N(s,a)}} \big\},
\end{eqnarray}
where $\hat r(s,a)$ is the sample average of (accumulated) returns by choosing action $a$ at state $s$, $N(s,a)$ is again the cumulative count number of choosing $a$ at state $s$ up to time $n$. 

In many scenarios like Grid Word or chess board games, the reward is very sparse. Therefore, the reward estimates $\hat r(s,a) \equiv 0$ before the agent can reach a non-zero reward state.
\eqref{eq:ucb} can be reduced to 
$\arg\max_a ~ \sqrt{\frac{\log n}{N(s,a)}}$ which is equivalent to 
$\arg\min_a ~ N(s,a)$.
The latter criterion is exactly the negative feedback algorithm.
Hence, by previous theorems, we know UCB algorithm is indeed theoretically better than naive random action selection in terms of exploration efficiency in very sparse reward environments.

\subsection{Connection with Monte Carlo Tree Search}

In computer science, Monte Carlo tree search (MCTS, \cite{browne2012survey, silver2016mastering}) is a heuristic search algorithm for some kinds of decision processes, most notably those adopted in software that plays board games. 
In that context, MCTS is usually used to solve the game tree.
MCTS consists of four main steps, selection, expansion, simulation and back propagation.

Recall the fact that, in expansion step, MCTS will always randomly choose the un-visited node rather than choose the visited nodes.
This exactly shares the same spirit as negative feedback algorithm does.
Moreover, in selection step, the agent usually chooses Upper Confidence Trees (UCT, \cite{kocsis2006bandit,browne2012survey, couetoux2011continuous}) criterion, 
$$\arg\max_a \frac{w_a}{n_{a}} + c \sqrt{\frac{\log N}{n_a}},$$ 
($n_a$ is the number of simulations after choosing action $a$; $w_a$ is the number of wins after choosing action $a$, $N$ is the number of simulations after the current node) to select the successive child nodes. 
If the tuning constant $c \rightarrow +\infty$, then UCT also reduces to $\arg\min_a n_a$ which is exactly how negative feedback algorithm chooses the next action.
Therefore, our new theories (partially) explains why combination of selection and expansion steps in MCTS is more efficient and effective than naive random tree search.

\subsection{Non-tabular Cases}

Up to now, we have only focused on graphs with finite nodes (i.e. discrete-state RL environments).
One may wonder can negative feedback algorithm be extended to non-tabular cases (i.e. the state space is continuous instead of being discrete)?
In below, we provide an approximate version of negative feedback algorithm without theoretical justification.

For arbitrary state $s$ and action $a$, we define the cumulative approximate visiting number as 
\begin{eqnarray}\label{proximal:visit}
N_{approx}^{(n)}(s,a) = \sum_{t' = 1}^{n-1} \kappa(s_{t'},s) \mathbf 1\{a_{t'} = a\}
\end{eqnarray} 
at time $n$.
Here kernel $\kappa(s_1, s_2)$ quantifies the similarity between two states. If states $s_1$ and $s_2$ are close, the value of $\kappa(s_1, s_2)$ will be close to 1. Otherwise, $\kappa(s_1, s_2)$ is close to 0. 
For example, in a Euclidean $\mathbb R^2$ space, a state $s$ can be represented by a two-dimensional coordinate, $(s_x, s_y)$.
The kernel function can be simply chosen as the indicator function, 
$\kappa(s_1, s_2) := \mathbf 1\{ |s_{1x} - s_{2x}| \leq \delta ~ \text{and}~ |s_{1y} - s_{2y}| \leq \delta\}$, where $\delta$ is a tuning parameter which adjusts the affinity level.
Then the agent will choose action $a$ in favor of the least approximate visiting number. That is,
\begin{eqnarray}
\pi_{approx}^{(n)}(s) = a ; ~ \text{if}~ a = \arg\min_{a'} N_{approx}^{(n)}(s,a'), 
\end{eqnarray}
where ties break randomly.

\section{Conclusion} \label{sec:conclusion}

In this work, we study 
the cover time problem of a non-Markovian algorithm, negative feedback strategy, which is based on "favor least" principle. 
To our knowledge, our work is the first theoretical work of this kind rather than empirical/synthetic study.
We make attempts to show that why negative feedback algorithm is better than naive random walk policy.
Specifically, we establish that the local version of negative feedback algorithm leads to a smaller expected number of excursions to visit any node in arbitrary graph.
We also establish that the cover time of negative feedback algorithm has smaller cover time under many special graphs, including star, clique graph, tree graphs, etc. 
Connections are made with several important RL algorithms, including maximum-entropy exploration, UCB and MCTS methods.
Various experimental results support our new theories and our findings. 
The result presented in this work may help practitioners to understand different exploration strategies better from mathematical angles. Theoretical analyses of cover time comparison in more complex graph structures or continuous-state environments can be considered as possible directions in future work.

%\nocite{langley00}

\bibliographystyle{plainnat}
\bibliography{reference}

\newpage
\clearpage

\appendix

\begin{center}
\large{
\textbf{Appendices of  "Exploration in Reinforcement Learning:\\
a Theoretical Study"}
}    
\end{center}

Experimental results, additional explanations and proofs of technical results are all collected in this appendix.

\section{Experimental Studies}
\label{sec:exp}

In this section, we numerically compare between negative feedback algorithm and random walk strategy under various RL environments. The detailed settings are given below.

\begin{figure}[ht]
	\centering
\includegraphics[width=0.24\textwidth]{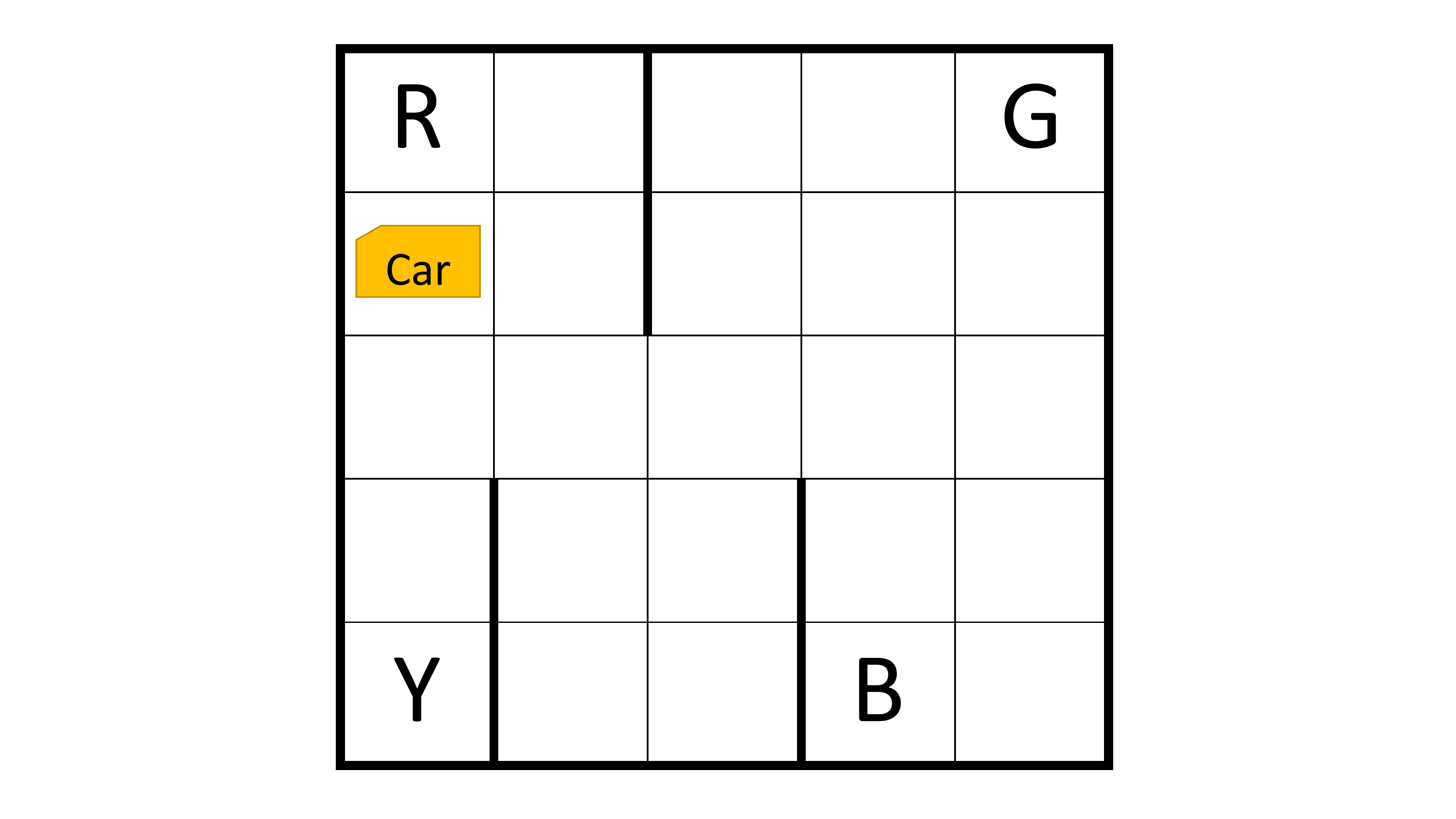}	\includegraphics[width=0.24\textwidth]{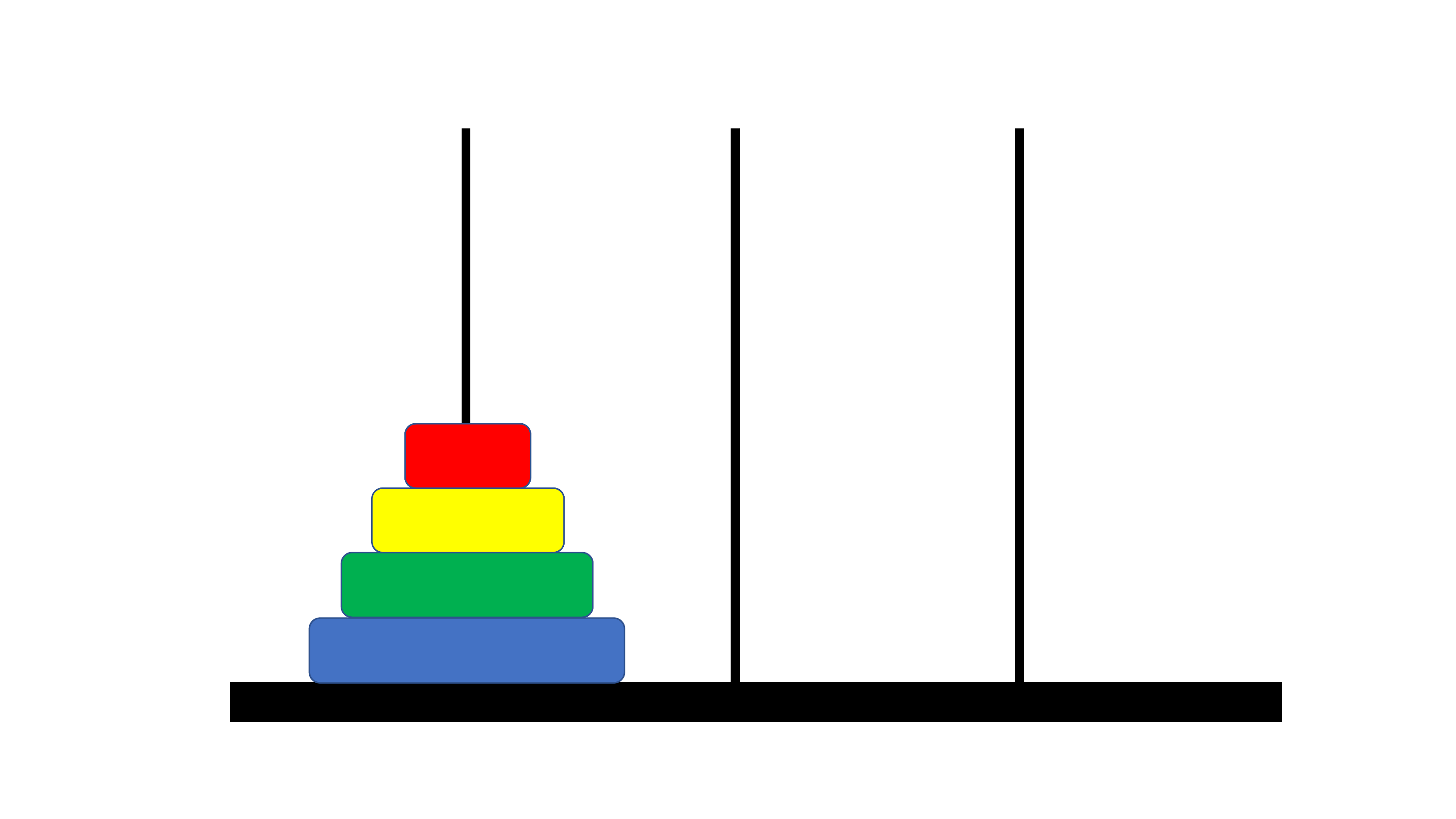} 
\includegraphics[width=0.24\textwidth]{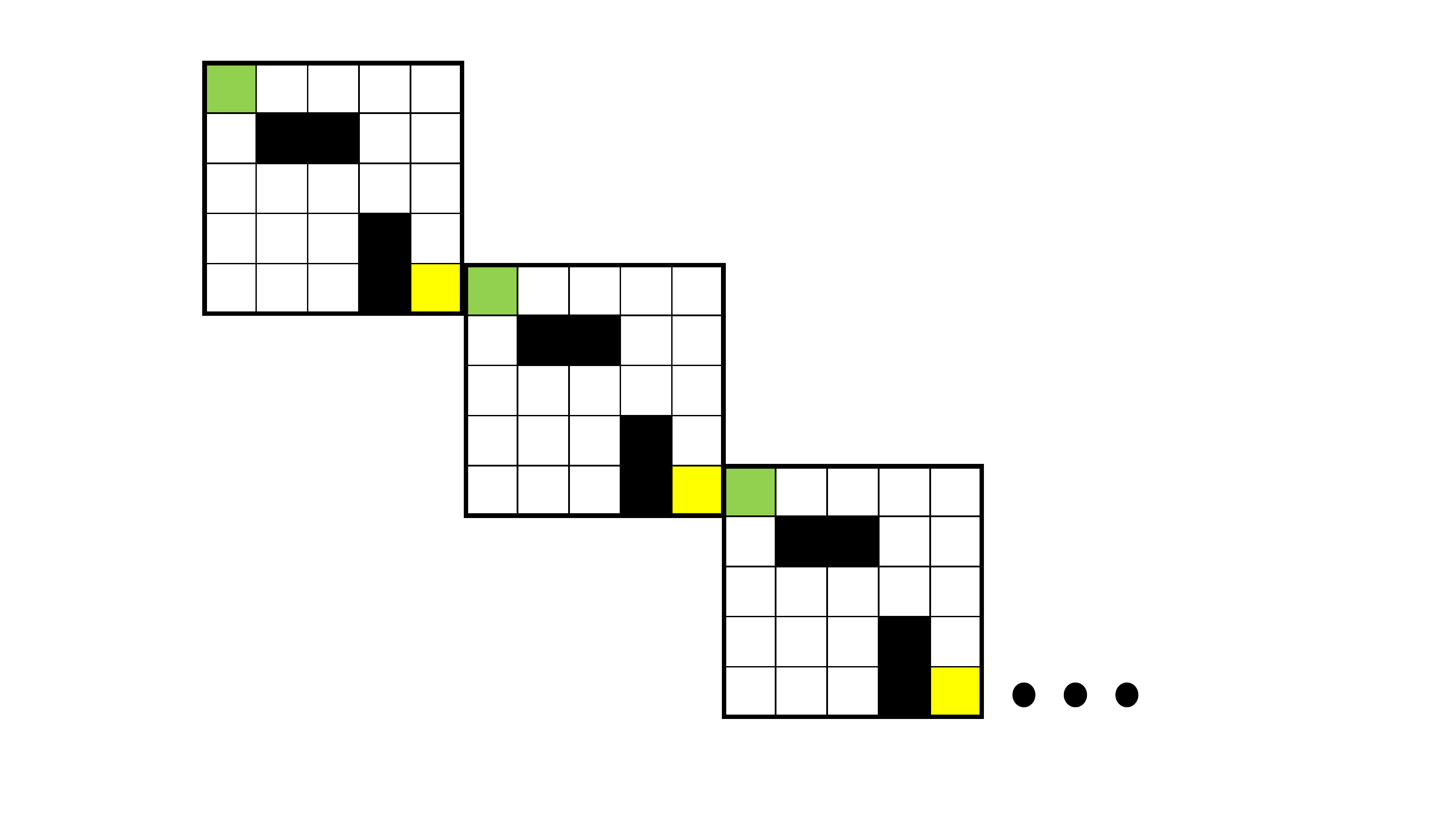} 
\includegraphics[width=0.24\textwidth]{images/tic-tac-toe.pdf} 
	\caption{First fig: taxi environment. Second fig: Hanoi environment. Third fig: multi-room environment. Fourth fig: tic-tac-toe game.
 }\label{fig:envs}
\end{figure}

\subsection{Settings}
\label{sec:exp1}

\noindent \textbf{Grid1Dim}: an environment with $n$ states lying on a line. One in state $i$ ($2 \leq i \leq n-1$) can move to either state $i-1$ and $i+1$. One can only move to 2 (or $n-1$) if it is in state 0 (or $n$).
The starting state is randomly chosen from 0 to $n$.

\noindent \textbf{GridCircle}: an environment with $n$ states lying forming a circle.
One in state $i$ ($1 \leq i \leq n$) can move to either state $i-1$ and $i+1$.
Here state -1 (or $n+1$) is treated as $n$ (or 0).
The starting state is again uniformly randomly chosen.

\noindent \textbf{Grid2dim}: A two-dimensional grid world with $n_1$ rows and $n_2$ columns.
At each state $(i,j)$ ($1 \leq i \leq n_1, 1 \leq j \leq n_2$), it can go up (down, left, right). That is, it transfers to next state $(i-1,j)$ ($(i+1,j)$, $(i,j-1)$, $(i,j+1)$).
If next state is out of boundary (i.e., $i+1 > n_1$, $i-1 < 1$, $j+1 > n_2$ or $j-1 < 1$), then it stays in the previous state.  
Again the starting state is uniformly randomly generated.
For simplicity, we set $n_1 = n_2 = grid \in \{5, 10, 20, 30, 40\}$.

\noindent \textbf{Grid3dim}:
A two-dimensional grid world with $n_1$ rows, $n_2$ columns and $n_3$ heights.
At each state $(i,j,k)$ ($1 \leq i \leq n_1, 1 \leq j \leq n_2, 1 \leq k \leq n_3$), it can go up (down, left, right, forward, back). That is, it transfers to next state $(i-1,j,k)$ ($(i+1,j,k)$, $(i,j-1,k)$, $(i,j+1,k)$, $(i,j,k-1)$, $(i,j,k+1)$).
If next state is out of boundary (i.e., $i+1 > n_1$, $i-1 < 1$, $j+1 > n_2$, $j-1 < 1$, $k+1 > n_3$ or $k-1 < 1$), then it stays in the previous state. 
The starting state is still uniformly randomly generated.
For illustration purposes, we set $n_1 = n_2 = n_3 = grid \in \{4, 6, 8, 10, 12\}$.

\noindent \textbf{Tree}: a tree-like graph. The depth is $n$ and each node has $m = 3$ children. Root node is numbered as $0$. For each non-leaf node $i$, it connects to children, $i\cdot m + 1, i \cdot m + 2, \ldots, (i+1) \cdot m$.
The root node is the starting state.

\noindent \textbf{BarBell}: The graph consists of two cliques with size $n$. The nodes in the first clique are numbered from 1 to $n$. The nodes from the second clique are numbered from $n+1$ to $2 n$. There is an extra link connecting node $n$ and $n+1$. 
The starting state is randomly chosen from 1 to $2n$.

\noindent \textbf{Taxi}: 
It is a 5 by 5 two dimensional grid space shown in the right of Figure \ref{fig:envs}.  The yellow icon represents a taxi car, which can move up (down, left, right). If it meets the borderline in boldface, then it remains the previous state. There are four target locations labeled with "R", "G", "Y", "B".
A passenger can randomly appear in one of these four positions. His/her destination is also a random place of these four locations.
The taxi needs to first pick up the passenger and then takes him/her to the destination.
At start, the car randomly appears in the 5 $\times$ 5 grid space and the passenger is waiting at "R", "G", "Y" or "B".

\noindent \textbf{Tower of Hanoi}:
This is a very classical game. There are four discs and three pegs. Initially, discs are placed in the leftmost peg as shown in middle of Figure \ref{fig:envs}. 
The game requires player to move all discs to the rightmost peg in the order that smaller disc is placed on the larger disc.
There is a rule that, in each move, it is only allowed to put a disc to an empty peg or onto a lager disc.

\noindent \textbf{MultiRoom}: It is a two-dimensional grid world such that $n$ identically same rooms are connected. The bottom-right grid of one room can transit to the top-left grid of next room. The black grids are not accessible.
In each state, it can go to one of four directions. It remains the previous state if it meets the borderline or black grids.
At start, the agent is placed at the most left and upper grid.

\noindent \textbf{Tic-tac-toe}: two players take turns marking the spaces in a three-by-three grid with cross (X) and nought (O).
One of two players who succeeds in placing three of their marks in a horizontal, vertical, or diagonal row is the winner.
Otherwise, the game ends with a draw.

\noindent \textbf{Continuous 2D Space}:
This is a $[0, D] \times [0, D]$ Eculidean space.
The agent randomly walks in either of $x$-axis or $y$-axis direction. 
Two walk types are considered. One is Brownian motion (the step size follows the standard Normal distribution). The other one is Levy flight \cite{viswanathan2000levy} (the step size follows Pareto distribution with degree 2). 
The whole space is equally divided into $M \times M$ sub-regions ( each sub-region is a square with side length of $\delta = D/M$).  
Cover time here is defined to be the expectation of first time of visiting all sub-regions.

\subsection{Results}

\begin{figure}
	\centering
\includegraphics[width=0.24\textwidth]{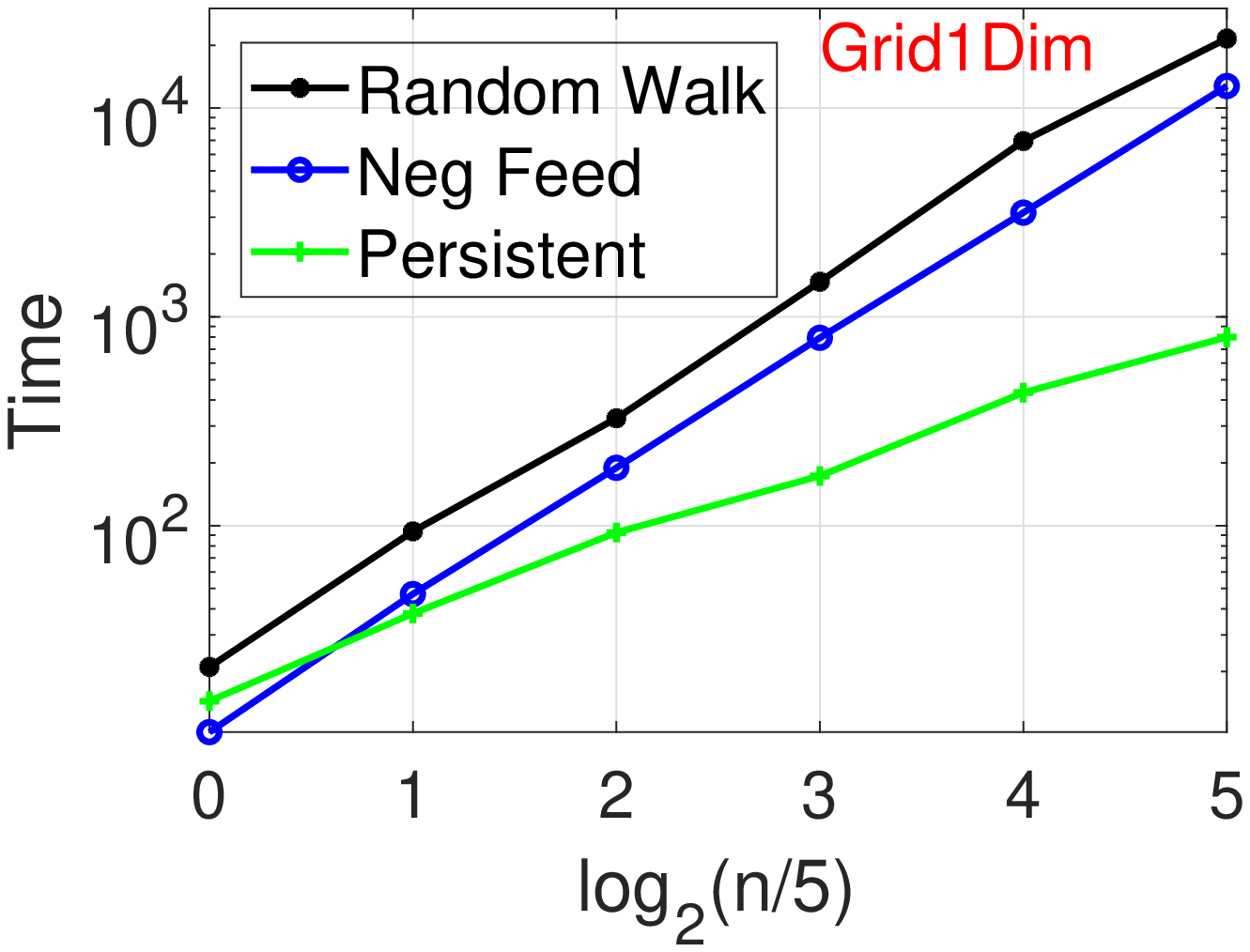}
\includegraphics[width=0.24\textwidth]{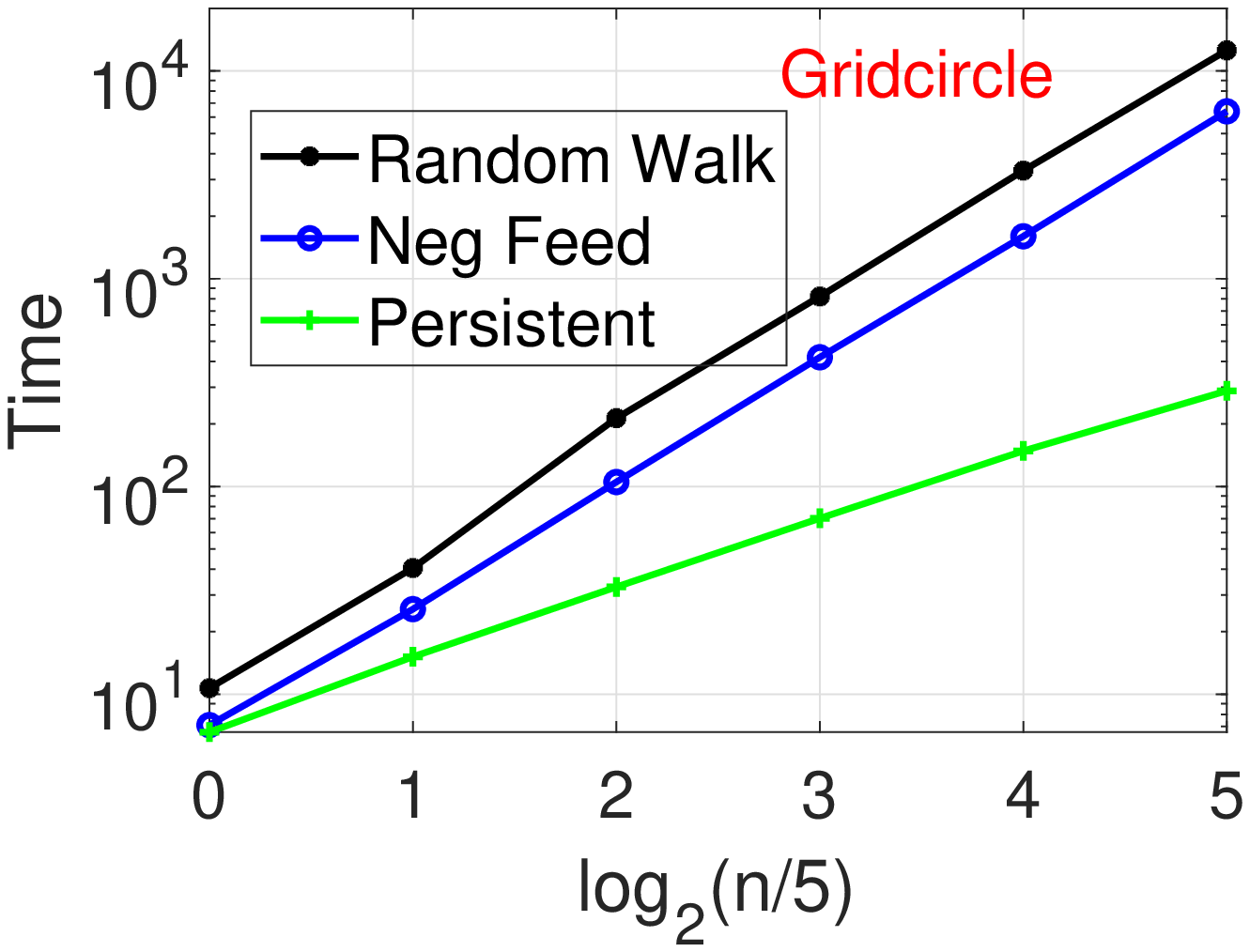} 
\includegraphics[width=0.24\textwidth]{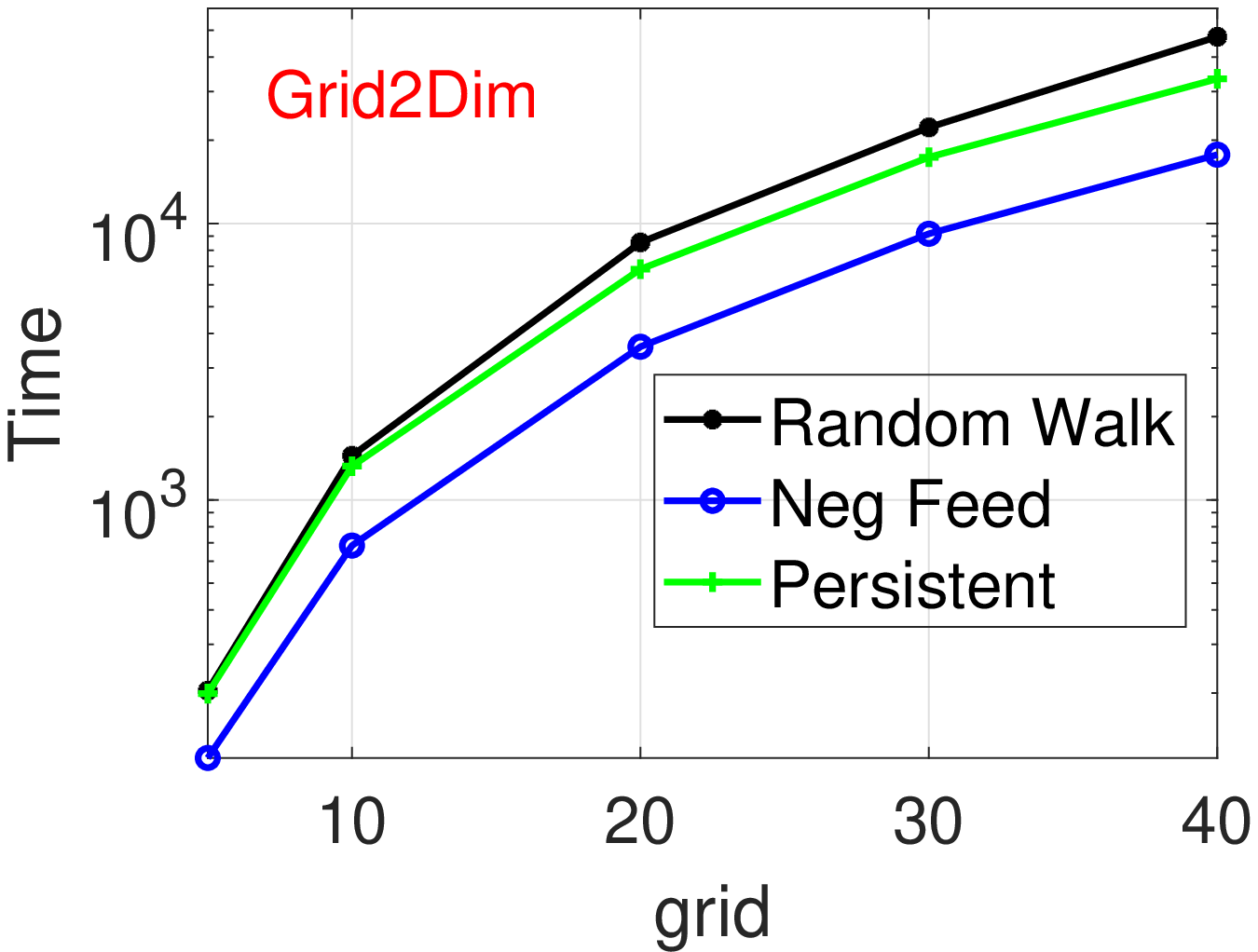} 
\includegraphics[width=0.24\textwidth]{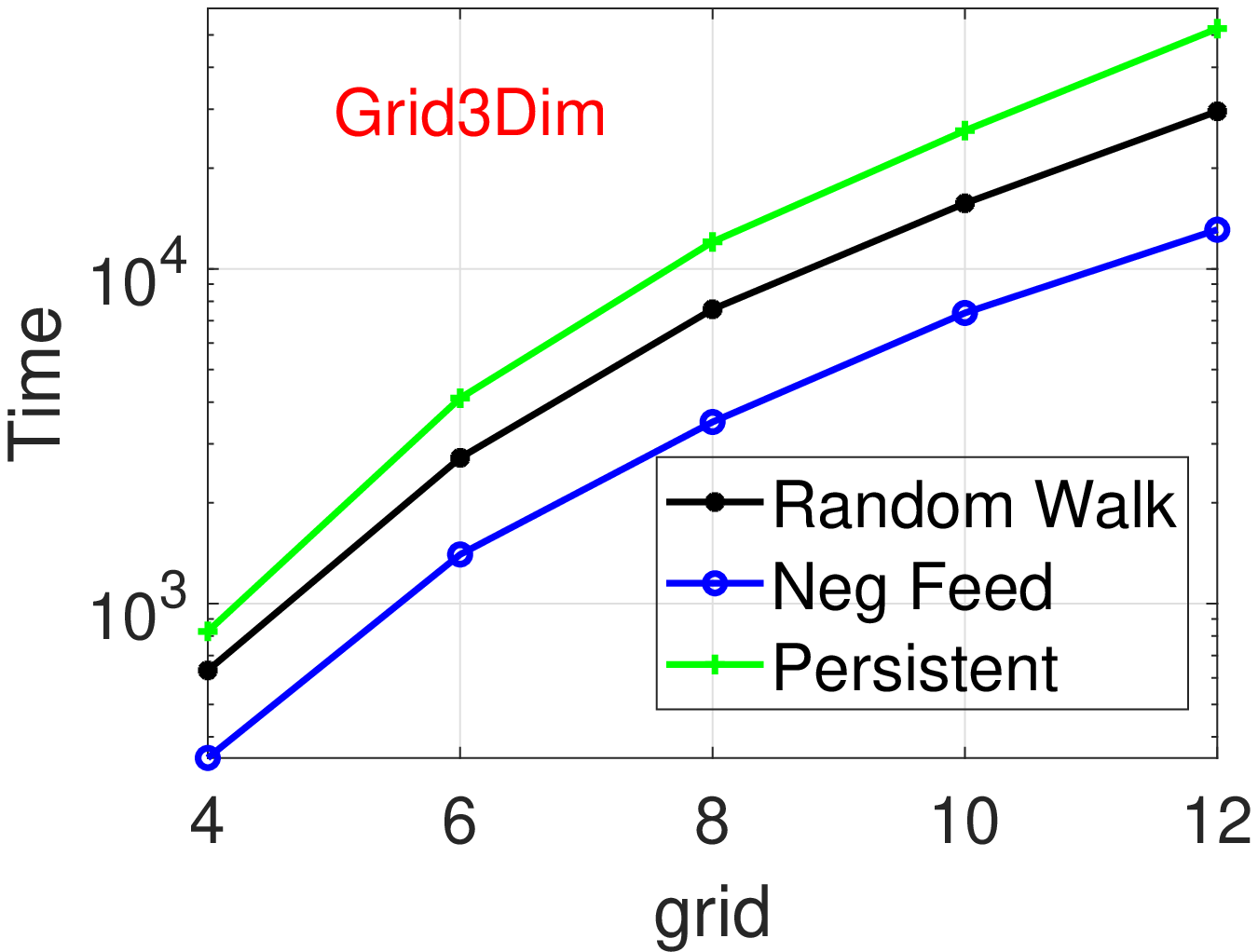} 
\includegraphics[width=0.24\textwidth]{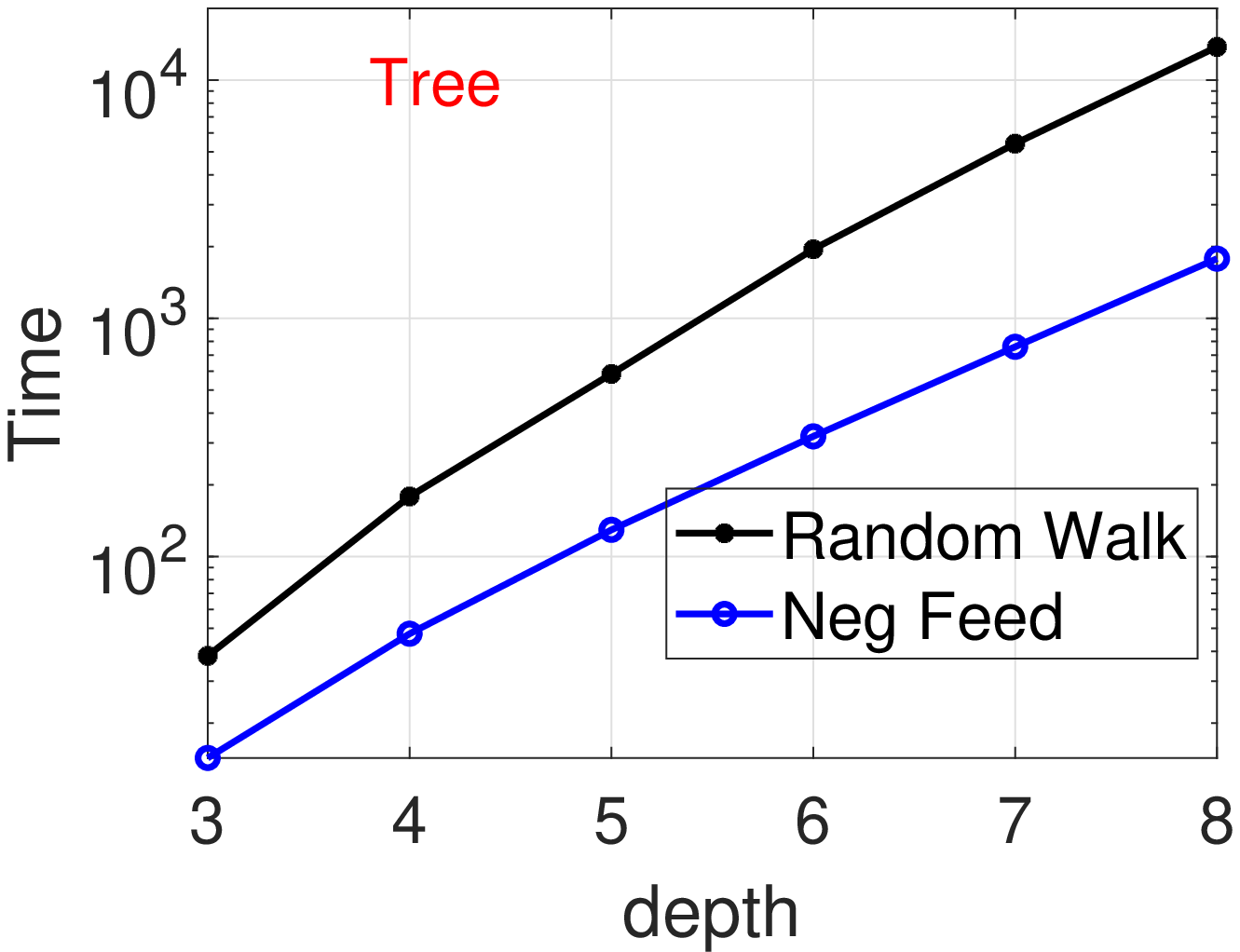} \includegraphics[width=0.24\textwidth]{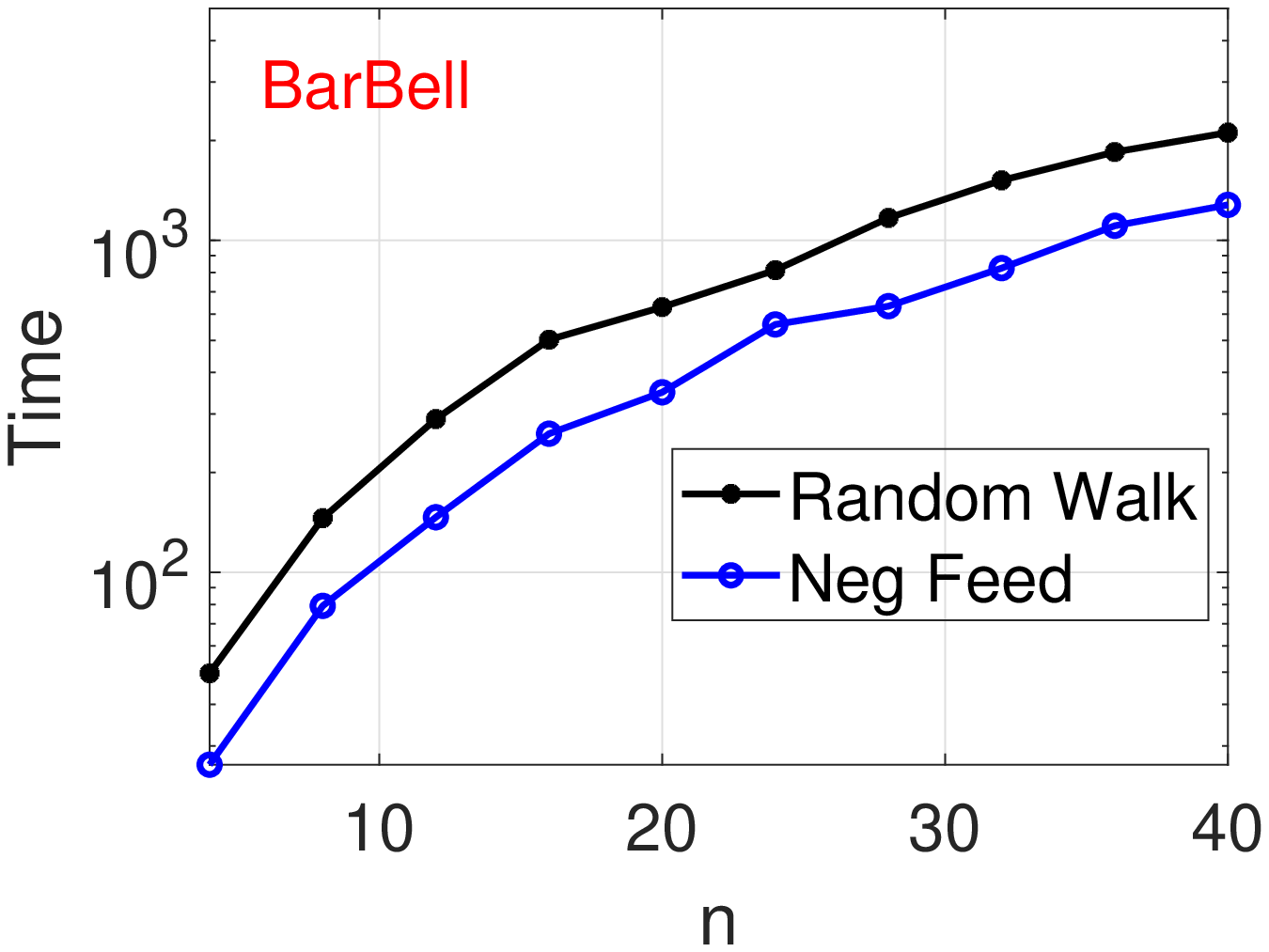}
\includegraphics[width=0.24\textwidth]{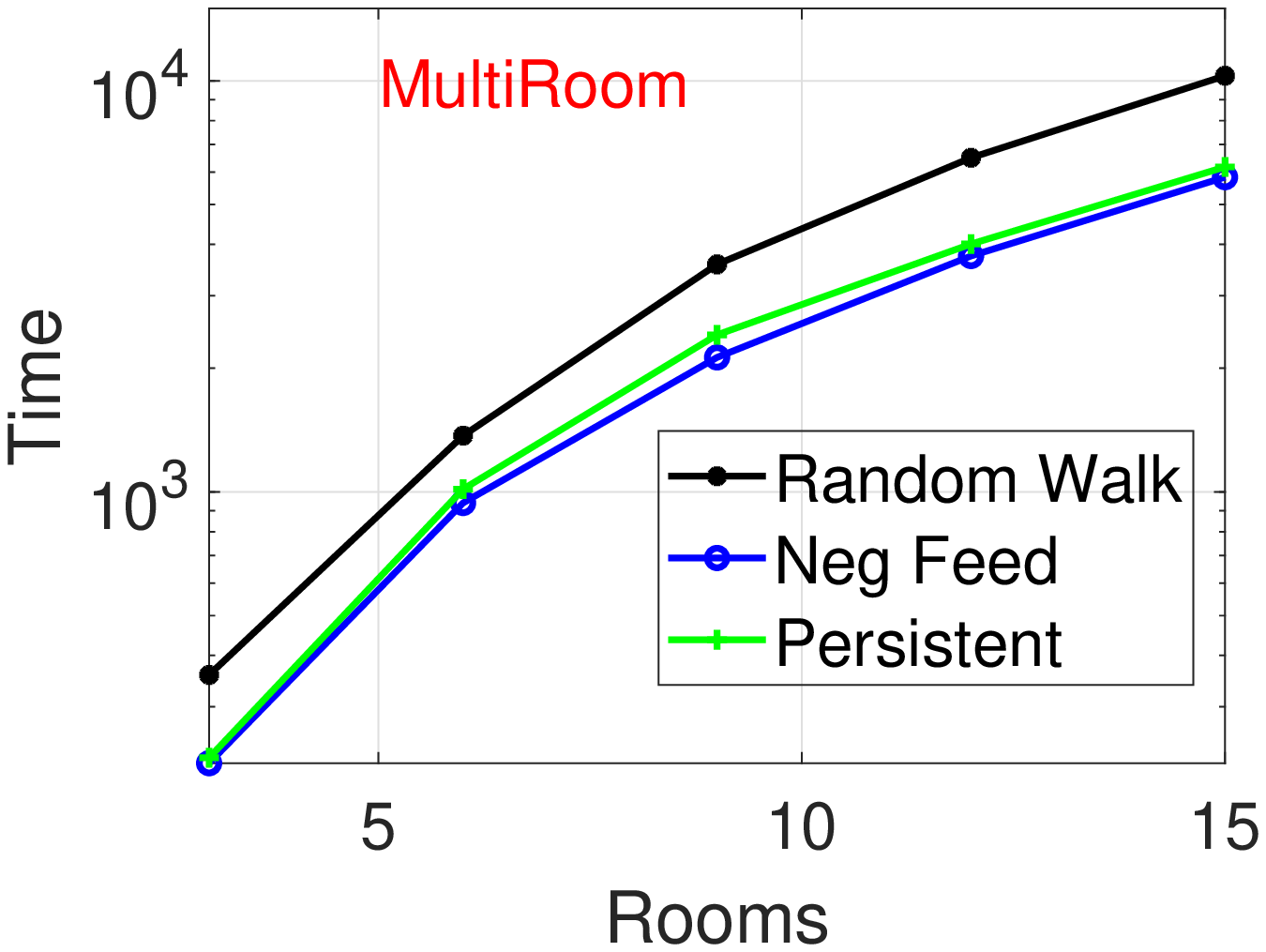} 
\includegraphics[height = 1in, width=0.23\textwidth]{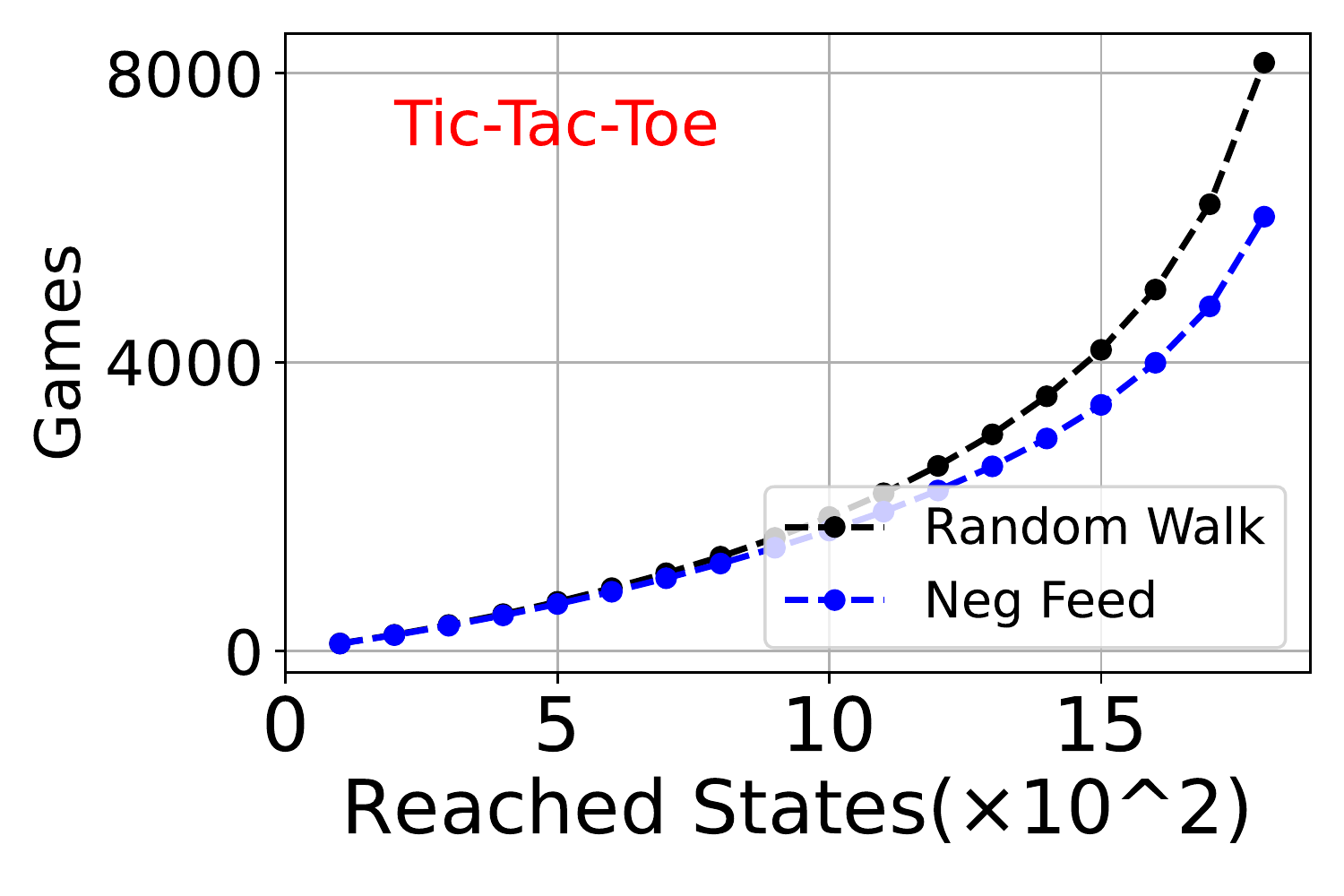}
\includegraphics[width=0.23\textwidth]{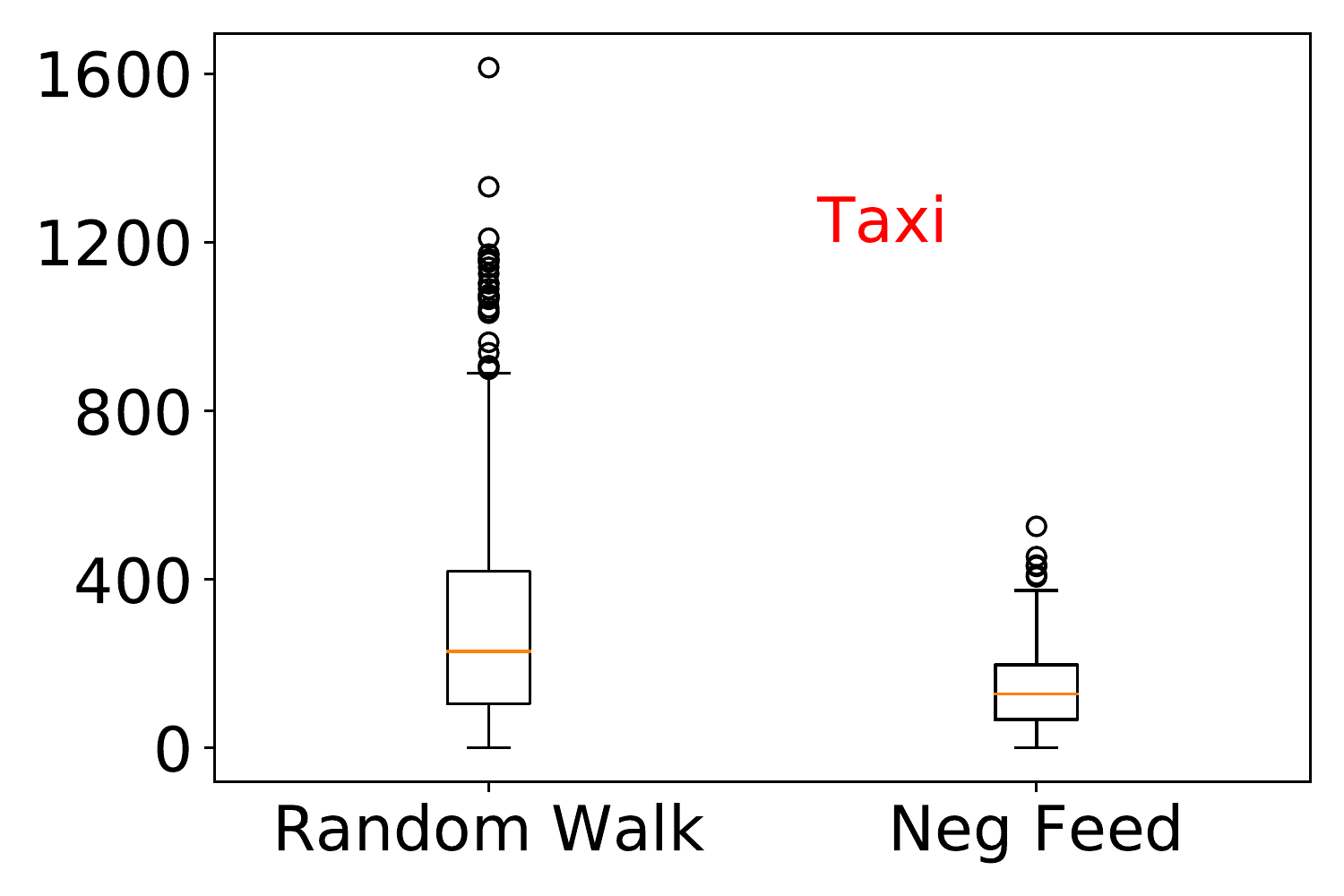}	
\hspace{3mm}
\includegraphics[width=0.23\textwidth]{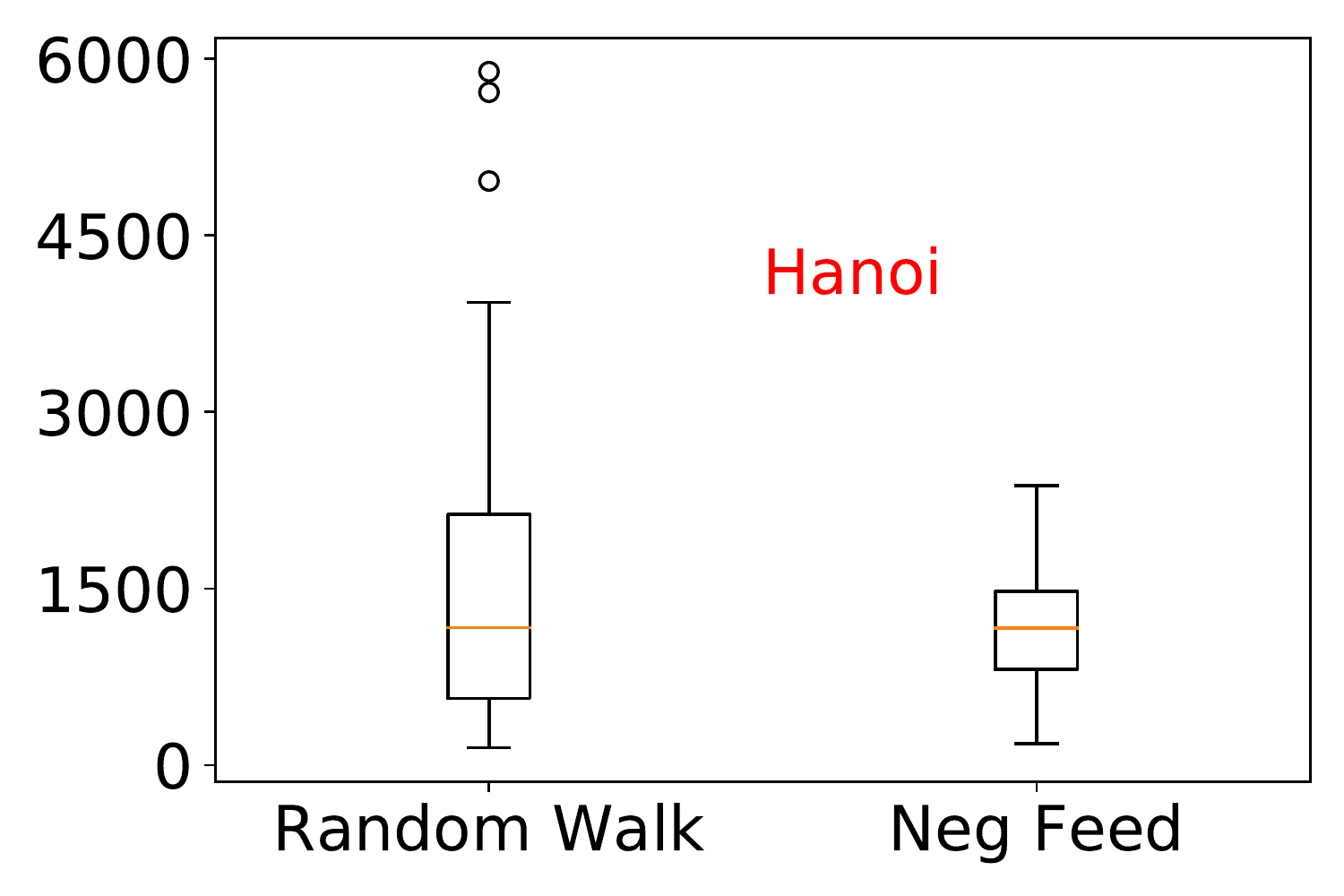} 
\includegraphics[width=0.23\textwidth]{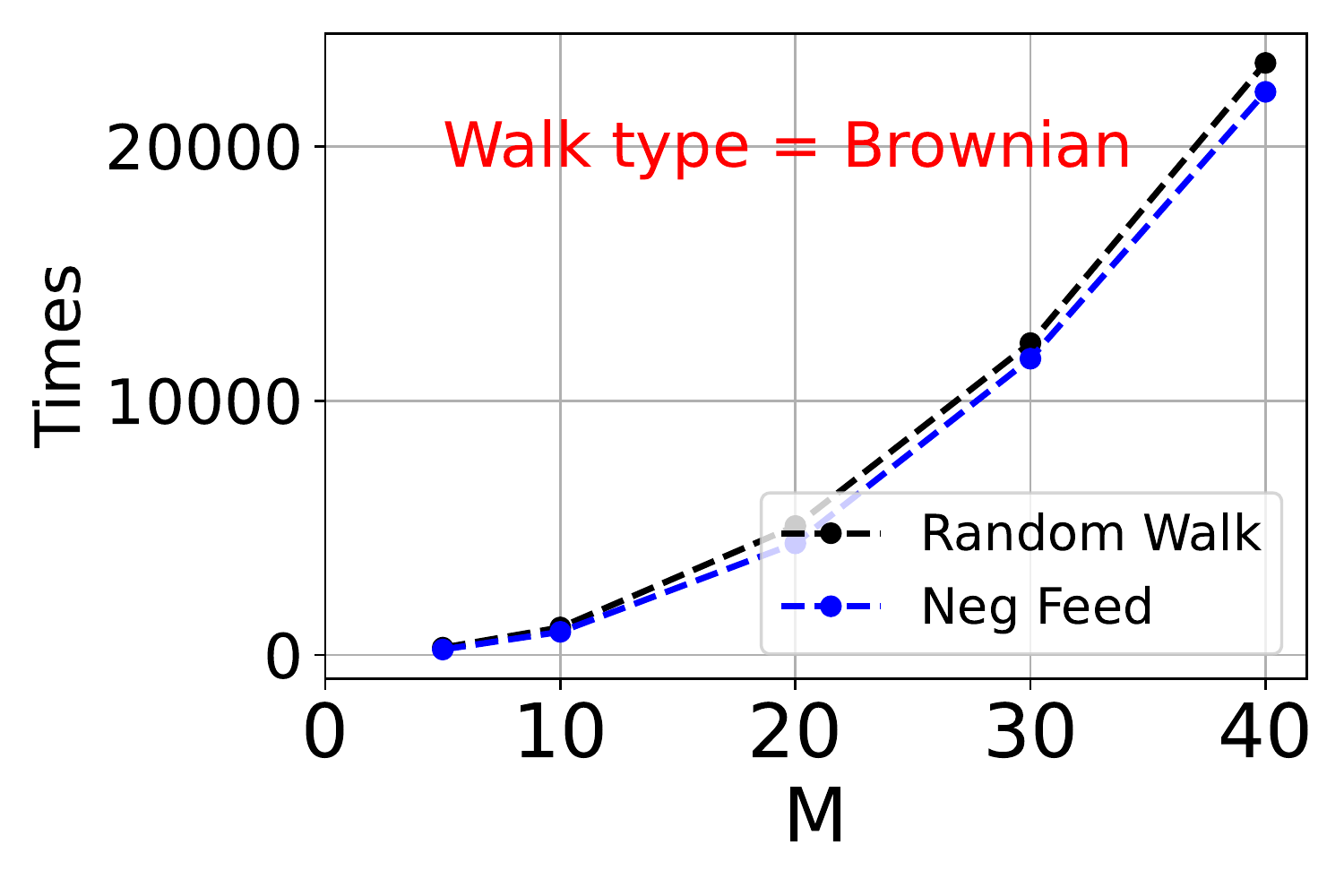} 
\includegraphics[width=0.23\textwidth]{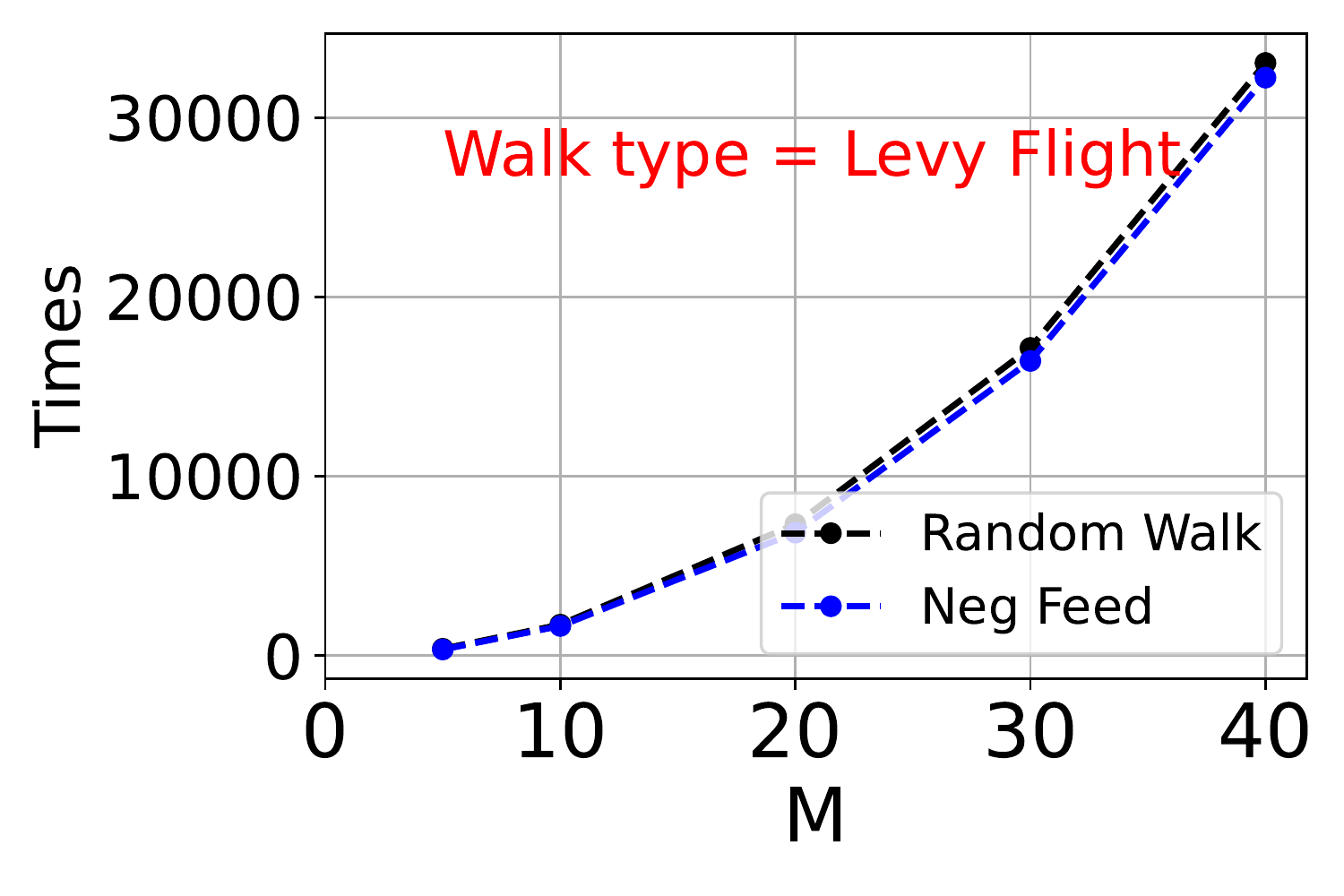}
\caption{Cover times under different policies in various discrete-state environments described in Section \ref{sec:exp1}.
Last two figures are for cover time comparison in 2D continuous-state environments. $D = 5, M \in \{5,10,20,30,40\}$.
}\label{fig:row1}
\end{figure}

For Grid World-type environments, we compare negative feedback algorithm, random walk algorithm and temporally extended/persistent algorithm (see definition in Section \ref{sec:2}) with $p(z) \propto 1/z$. 
From Figure \ref{fig:row1}, we can see that negative feedback is universally better than random walk strategy. Temporally persistent method has smaller cover time in 1-dimensional environment but becomes worse in 2 and 3-dimensional cases.

For non Grid World-type environments, we only compare negative feedback and random walk algorithms without considering temporally persistent strategy. This is because the latter method makes agent easily get stuck at certain states and thus delays the whole exploration procedure. 
In Tree graph, we can see that the negative feedback strategy improves cover time significantly over random walk strategy as height of tree goes deep.
In the tasks of Taxi and Tower of Hanoi environments, the negative feedback algorithm also uses much less time to reach the target state compared with that of random walk policy.
\ref{sec:4}.
In Tic-tac-toe game, we plot the curves of "number of games played" versus "number of visited end states". (An "end state" is a leaf-node at which the game ends.)
We can see that negative feedback algorithm takes much fewer games to explore the same number of end states. In other words, compared to random walk strategy, negative feedback method can learn all possible final situations of tic-tac-toe game much faster. 
All these phenomena corroborate our theories developed in Section \ref{sec:4}.

Lastly, in two-dimensional continuous space search problems, we can see that the negative feedback algorithm also improves cover time over the random walk policy from Figure \ref{fig:row1}.
This sheds a light on possible future work on theoretical comparison between different policies in non-discrete RL problems.

\section{More Detailed Connections to RL Problems} 

In the reinforcement learning (RL) problem,
an agent must interact with the environment, by taking actions and observing their consequences.
When the agent starts acting in an environment, it usually does not have any prior knowledge regarding the task which it needs to tackle \citep{amin2021survey}.
The ability of the agent to explore the environment fast is a key factor in success of achieving higher cumulative rewards. 
Efficient exploration has been acknowledged as an important area in adaptive control
since decades ago, starting with the multi-armed bandit problems \citep{thompson1933likelihood, lattimore2020bandit}. It has also become a hot topic in RL literature \citep{sutton1995generalization} since 1990s.  

The categorization of exploration methods in RL can be described as in Table \ref{tab:cat}.
The two main categories are "Reward-free" approaches and "Reward-based" approaches.
In the former category, the methods do not take into account reward in their action-selection criteria.
By contrast, in the latter one, the methods select the action based on the rewards returned by the environment.

\begin{table}[ht]
    \centering
    \begin{tabular}{c|c|l}
    \hline
    \hline
    \multirow{5}{*}{Category}   & \multirow{2}{*}{Reward-free}  & {\color{black} Blind} \\
       &  & Intrinsically Motivated \\
    \cline{2-3}
    & \multirow{3}{*}{Reward-based} & Randomized action selection \\
    & & Probability matching\\
    & & Optimism Based \\
    \hline
    \hline
    \end{tabular}
    \caption{The categorization of exploration strategies in RL. Main focus of this work is on reward-free strategies.}
    \label{tab:cat}
\end{table}

Moreover, there are two sub-categories in "Reward-free" strategies, "Blind" exploration and "Intrinsically-motivated" exploration.
"Blind" methods explore the environment solely based on random action selection. The agent is not guided by the past information history.
Such technique is the most basic, simple and effective type of methods in the literature. Its special cases include random-walk \citep{thrun1992efficient}, 
$\epsilon$-greedy method \citep{caironi1994training, sutton1995generalization},
temporally extended $\epsilon z$-greedy method \citep{dabney2020temporally}, etc.
In contrast to blind exploration, intrinsically-motivated exploration methods utilize some environment intrinsic structure to encourage exploring the entire space.
Many of such methods aim at minimizing agent's prediction uncertainty or error \citep{schmidhuber1991curious,schmidhuber1991possibility,pathak2017curiosity}. 
Other of such methods aim to pursue the maximization of space coverage \cite{machado2017laplacian, machado2017eigenoption, hong2018diversity, jinnai2019discovering, jinnai2019exploration}.

On the other hand, there are three sub-categories in "Reward-based" strategies, "Randomized action selection" exploration, "Probability matching" exploration and "Optimism (bonus) Based" exploration. 
Specifically, "Randomized action selection" 
assigns action selection probabilities to the admissible actions based on value functions/rewards (e.g. Boltzmann distribution \citep{watkins1989learning, lin1992self}, value-difference based exploration \citep{tokic2010adaptive, tokic2011value}) or the learned policies (e.g. policy gradient \citep{kakade2001natural, silver2014deterministic}).
Probability matching is also known as Thompson sampling \citep{thompson1933likelihood}, which maintains a posterior distribution over its beliefs regarding the optimal
action, instead of directly selecting the action with the highest expected return.
Posterior Sampling for Reinforcement Learning (PSRL, \cite{strens2000bayesian}) and its variants \citep{osband2013more, osband2019deep} are shown to be successful in many RL tasks.
In the field, one of most popular sub-categories of exploration methods is the optimism (bonus)-based method \citep{kaelbling1996reinforcement}, where
algorithm employs different bonus calculation technique to encourage the choice of action that leads to a higher level of uncertainty and consequently, novel or informative states.
Upper Confidence Bounds (UCB, \cite{strehl2006pac, auer2006logarithmic, jaksch2010near}), Upper-Confidence Least-Squares (UCLS, \cite{kumaraswamy2018context}), $E^3$ \citep{kearns2002near} belong to this domain.

% In reinforcement learning literature, one important and popular type of strategy is called $\epsilon$-greedy exploration.
% The core idea is that 
% one performs the best possible action according to the learned
% policy with $1-\epsilon$ probability and chooses an action \textbf{completely at random} with a probability of $\epsilon$.

In reinforcement learning problems, a common phenomenon is that the environment provides very sparse rewards, that is, the agent can observe (positive) rewards only if it reaches a very few specific target states. 
For example, in Grid World game \citep{stolle2002learning, tizhoosh2005reinforcement}, the robot moves up, down, left or right in a grid board. It can only be fed with positive feedback once it gets the target grid. 
In board game like Tic-tac-toe \citep{beck2008combinatorial, abu2019tic}, two players take turns marking the spaces in a three-by-three grid with cross (X) and nought (O).
The reward will not be given only until one player wins or a draw happens.
Therefore, fast exploration of the entire environment space without knowledge of extrinsic rewards plays a key role in success of a RL agent / strategy.
An interesting research question naturally arises that, 
\textit{with reasonable theoretical guarantees}, could we improve "blind" method by using \textbf{reward-free} strategy and choosing action a bit more smartly rather than taking action completely at random (i.e. random walk)?
To be more mathematically formal, we want to find an exploration strategy whose cover time (i.e. the expectation of first time of covering all possible states in the environment) is \textbf{smaller} than that of pure random search under certain environment assumptions or structures. 

To our knowledge, there is no literature directly related to cover time comparison between non-trivial exploration strategy and pure random walk strategy. 
Despite this, some strategies among blind exploration category touch upon cover time minimization. 
\cite{dabney2020temporally} introduces a temporally-extended $\epsilon$-greedy method, which utilizes the option (a strict generalization of action) strategy allowing repeated actions for consecutive times. They empirically show the advantage over random search in Deep Sea, Mountain Car environments.
\cite{jinnai2019discovering, jinnai2019exploration} compute the Laplacian matrix and try to maximize the second smallest eigenvalue (known as algebraic connectivity \citep{fiedler1973algebraic}). Their methodology is heuristic and the construction of Laplacian requires sufficient large action-state samples.

As a result, the problem considered in current work is fundamental. Our results (partially) address the issue that why "favor least" can perform much more efficiently than random walk from cover-time viewpoint.

\section{Proof of Results in Section \ref{sec:2}.}

\begin{proof}[Proof of Proposition \ref{prop:2}]
We first code 7 states for this toy grid world as below, "Start":0, upper-left grid: 1, bottom-left grid: 2, middle grid: 3, middle-right grid: 4, upper-right grid: 5, and "End":6.
We next define $T_{i}$ ($i \in \{0,1,2,3,4,5\}$) as the expectation of first time to reach state 6 by starting at state $i$.

Next We let $a$ be the probability of performing an action for 1 consecutive time (i.e. $p(z = 1) = a$) and $b$ be the probability of performing an action for 2 consecutive times (i.e. $p(z = 2) =  b = 1 - a$).
There is no need to consider performing an action for more than 2 consecutive times, since the maze is only of size, three by three. (If an agent repeat the same action for $\geq 3$ times, then it will hit the wall and waste the time.)

By directly examining the specific structure of this grid world, we can get the following recursive formula.
\begin{eqnarray}
T_0 & = & \frac{1}{3} T_1 + \frac{1}{3} T_2 + \frac{a}{3} T_3 + \frac{b}{3} T_4 + a + 2b, \label{T0} \\
T_1 & = & a T_0 + b T_2 + a + 2b, \nonumber \\
T_2 & = & a T_0 + b T_1 + a + 2b, \nonumber \\
T_3 & = & \frac{1}{2} T_0 + \frac{1}{2} T_4 + a + 2b, \nonumber \\
T_4 & = & \frac{1}{3} T_5 + \frac{a}{3} T_3 + \frac{b}{3} T_0 + 2 (a + 2b)/3 + 1/3, \nonumber \\
T_5 & = & a T_4 + a + 2b. \nonumber
\end{eqnarray}
By simplifications, we know 
$$T_1 = T_2 = \frac{a}{1 - b} T_0 + \frac{a + 2b}{1 - b};$$
$$(1 - \frac{a}{3}) T_4 = \frac{a}{3} T_3 + \frac{b}{3} T_0 + (a + 2b) + \frac{1}{3};$$
$$T_3 = \frac{1}{2}(T_0 + T_4) + (a + 2b).$$
Furthermore, we have
\begin{eqnarray}
(1 - a/2) T_4 &=& (a/6 + b/3) T_0 + (a/3 + 1)(a + 2b) + \frac{1}{3}, \\ 
(2 - a) T_3 &=& (1 - a/3 + b / 3) T_0 + (3 - 2 a/3)(a + 2b) + \frac{1}{3}.
\end{eqnarray}
Solving above equations for $T_3$, $T_4$ and plugging back to \eqref{T0}, we get
\begin{eqnarray}
T_0 = \frac{\frac{2 (a + 2b)}{3(1-b)} + (a + 2b)\frac{3a - 2a^2/3 + 2ab/3 + 2b + 1/3}{3(2-a)} + a + 2b}{
1 -  \frac{2 a }{3 (1 - b)} - \frac{a - a^2/3 + 2 ab /3 + 2 b^2/3}{3(2-a)} }. \label{a-b}
\end{eqnarray}
By noticing that \eqref{a-b} is a decreasing function of variable $a$.
Therefore the smallest $T_0$ is 23 when $a = 1$ and $b = 0$.
This implies that the random walk policy is always better than temporally persistent policy in this environment.
\end{proof}

\begin{proof}[Proof of Proposition \ref{prop:1}]
For negative-feedback algorithm, we first consider a restricted environment that state 1 and 2 are not accessible. Then there are only five reachable states, i.e., 0,3,4,5,6.
Under this restricted environment, we can compute the $T_0'$ by listing all possible paths in the following table.

Based on Table \ref{tab:paths}, we can easily compute that
$T_0' = 95/12$ in the restricted environment.
Moreover, we can see that the policy goes back to state 0 at most 2 times (excluding the initial position).
In other words, in original environment, state 1 and 2 can be visited at most three times.  
Therefore, $\mathbb E[T_{start, right}] = 95/12$, which gives $T_0 \leq T_0' + (2 + 2) * 3 < 23$. This proves that the negative feedback algorithm is better than random walk algorithm in this task. 

\begin{table}[ht!]
	\centering
	\begin{tabular}{ c|c|c } 
		\hline
		\hline
		& Path & Probability  \\
		\hline 
		1& (0,3,5,6) & 1/6 \\
		2& (0,3,4,5,4,6) &  1/12 \\
		3& (0,3,4,5,4,3,0,3,4,6) & 1/12 \\
		4& (0,3,0,3,4,6) & 1/6 \\
		5& (0,3,0,3,4,5,4,6) &  1/12 \\ 
		6& (0,3,0,3,4,5,4,3,4,6) & 1/24 \\
		7& (0,3,0,3,4,5,4,3,0,3,4,6) & 1/24 \\
		8& (0,3,4,3,0,3,4,6) & 1/24 \\
		9& (0,3,4,3,0,3,4,5,4,6) & 1/24 \\ 
		10& (0,3,4,3,0,3,0,4,4,6) & 1/24 \\
		11& (0,3,4,3,0,3,0,3,4,5,4,6) & 1/24 \\
		12& (0,3,0,3,4,3,4,6) & 1/24 \\
		13& (0,3,0,3,4,3,4,5,4,6) & 1/24 \\
		14& (0,3,0,3,4,3,0,3,4,6) & 1/24 \\
		15& (0,3,0,3,4,3,0,3,4,5,4,6) & 1/24 \\
		\hline
	\end{tabular}
\caption{In the restricted 5-state grid world, all possible paths along with corresponding probability under negative feedback policy.}\label{tab:paths}
\end{table} 
\end{proof}

\section{Proof of Result in Section \ref{sec:3}}

\begin{proof}[Proof of Theorem \ref{thm:local}]
Suppose that node $i$ has $K$ neighbours in the graph, denoted as $a_1,\ldots,
a_K$, so that $d_i=K$. 
For $k=1,\ldots, K$, we define
\begin{equation} \label{e:success.p}
p_{a_k} = \mathbb P\bigl( T_j\leq V_1|X_0=i,X_1=a_k\bigr),
\end{equation}
the probability of visiting node $j$ before returning to node
$i$. Then a simple recursive argument shows that under the random walk
dynamics,
\begin{align} \label{e:RW.rec}
\mathbb E_{\pi_{rw}}[N_j|X_0=i] =& \frac{1}{K} \sum_{k=1}^K \bigl[ p_{a_k}\cdot 1+
(1-p_{a_k})\bigl(1+ \mathbb E_{\pi_{rw}}[N_j|X_0=i] \bigr)\bigr] \\
\notag =& 1+ \mathbb E_{\pi_{rw}}[N_j|X_0=i] \cdot \frac{1}{K} \sum_{k=1}^K  (1-p_{a_k}) \\
\notag =& 1+ \mathbb E_{\pi_{rw}}[N_j|X_0=i] \left( 1-\frac{1}{K} \sum_{k=1}^K
p_{a_k}\right). 
\end{align}
This means that,  under the random walk dynamics,
\begin{align} \label{e:RW.fin}
\mathbb E_{\pi_{rw}}[N_j|X_0=i] =\frac{K}{\sum_{k=1}^K  p_{a_k}}.
\end{align}

\bigskip

The modified local negative feedback algorithm is not Markovian, but there
exist regeneration points as well. In fact, these are the time points when the agent
returns back to node $i$ and by that time the process has moved from state
$i$ to each one of its $K$ neighbours with equal number of times (i.e. those time points $n$'s such that $Kmin_i^{(n)} = d_i$). Beginning
at each such a regeneration point, the sequence of the next $K$
actions (steps out
of the node $i$) will be a random permutation $\pi = (\pi_1,\ldots,
\pi_k)$ of the action set $\{a_1,\ldots, a_K\}$, after which a
regeneration point is reached once again. Therefore, a recursive
argument shows that
\begin{align} \label{e:NF.rec}
\mathbb E_{\pi_{loc}}[N_j|X_0=i] =& \sum_{\pi = (\pi_1,\ldots,
	\pi_k)} \sum_{k=1}^K \frac{1}{K}\frac{1}{K-1}\cdots \frac{1}{K-k+1}
\prod_{m=1}^{k-1}(1-p_{a_{\pi_m}})  p_{a_{\pi_k}} \cdot k \\
\notag    +&    \prod_{k=1}^{K}(1-p_{a_k})\bigl( K+  \mathbb E_{\pi_{loc}}[N_j|X_0=i]\bigr).
\end{align}
We can reorganize the first term in the right hand side of
\eqref{e:NF.rec} by collecting together all terms in the sum which
correspond to the same $k=1,\ldots, K$.  This puts the first term in
the right  hand side of \eqref{e:NF.rec}  in the form 
\begin{align*}
&\sum_{k=1}^K k\frac{(k-1)!}{K(K-1)\cdots (K-k+1)}
\sum_{|J|=k-1} \sum_{l\not\in J} p_{a_l}\prod_{j\in J}(1-p_{a_j}) \\
= &\sum_{k=1}^K   \frac{1}{{K \choose k}}
\sum_{|J|=k-1} \sum_{l\not\in J}
\bigl(1-(1-p_{a_l})\bigr)\prod_{j\in J}(1-p_{a_j}) \\
=&\sum_{k=1}^K   \frac{1}{{K \choose k}} (K-k+1)
\sum_{|J|=k-1}\prod_{j\in J}(1-p_{a_j})
- \sum_{k=1}^K   \frac{1}{{K \choose k}} k 
\sum_{|J|=k}\prod_{j\in J}(1-p_{a_j})\\
= &\sum_{k=0}^{K-1}   \frac{1}{{K \choose k+1}} (K-k)
\sum_{|J|=k}\prod_{j\in J}(1-p_{a_j})
- \sum_{k=1}^K   \frac{1}{{K \choose k}} k 
\sum_{|J|=k}\prod_{j\in J}(1-p_{a_j}) \\
=& 1- K\prod_{j=1}^K (1-p_{a_j})
+ \sum_{k=1}^{K-1} \left( \frac{(K-k)(k+1)!(K-k-1)!}{K!}-
\frac{kk!(K-k)!}{K!}\right)  \sum_{|J|=k}\prod_{j\in
	J}(1-p_{a_j}) \\
=&1- K\prod_{j=1}^K (1-p_{a_j})
+ \sum_{k=1}^{K-1} \frac{1}{{K \choose k}} \sum_{|J|=k}\prod_{j\in
	J}(1-p_{a_j}) . 
\end{align*}
Solving now \eqref{e:NF.rec} says that under the modified local negative
feedback algorithm, 
\begin{align} \label{e:NF.fin}
\mathbb E_{\pi_{loc}}[N_j|X_0=i] =\frac{1+\sum_{k=1}^{K-1} \frac{1}{{K \choose k}} \sum_{|J|=k}\prod_{j\in
		J}(1-p_{a_j}) }{1-\prod_{j=1}^K (1-p_{a_j})}.
\end{align}

Our goal is to show that the expected value \eqref{e:NF.fin}  
under the modified negative  feedback dynamics is never larger than 
the expected value \eqref{e:RW.fin}  under the random walk dynamics. 
That is, we need to show that
\begin{equation} \label{e:comp}
\frac{K}{\sum_{j=1}^K  p_{a_j}} \geq 
\frac{1+\sum_{k=1}^{K-1} \frac{1}{{K \choose k}}
	\sum_{|J|=k}\prod_{j\in 
		J}(1-p_{a_j}) }{1-\prod_{j=1}^K (1-p_{a_j})}.
\end{equation}
This is, of course, the same as
\begin{equation} \label{e:comp1}
1-\frac{\sum_{j=1}^K (1- p_{a_j})}{K}\leq
\frac {1-\prod_{j=1}^K (1-p_{a_j})}{1+\sum_{k=1}^{K-1} \frac{1}{{K
			\choose k}}   \sum_{|J|=k}\prod_{j\in 
		J}(1-p_{a_j}) } .
\end{equation}
Denoting $x_j=1-p_{a_j}, \, j=1,\ldots, K$, we need to prove that
\begin{equation} \label{e:comp2}
1-\frac{\sum_{j=1}^K  x_j}{K}\leq
\frac {1-\prod_{j=1}^K  x_j}{1+\sum_{k=1}^{K-1} \frac{1}{{K
			\choose k}}   \sum_{|J|=k}\prod_{j\in 
		J} x_j } .
\end{equation}
Alternatively, we need to prove that 
\begin{align} \label{e:comp3} 
\frac{\sum_{j=1}^K  x_j}{K}\geq& 
\frac{\sum_{k=1}^{K-1} \frac{1}{{K
			\choose k}}   \sum_{|J|=k}\prod_{j\in 
		J} x_j + \prod_{j=1}^K  x_j}{1+\sum_{k=1}^{K-1} \frac{1}{{K
			\choose k}}   \sum_{|J|=k}\prod_{j\in 
		J} x_j }\\
\notag =&\frac{\sum_{k=1}^{K} \frac{1}{{K
			\choose k}}   \sum_{|J|=k}\prod_{j\in 
		J} x_j}{\sum_{k=0}^{K-1} \frac{1}{{K
			\choose k}}   \sum_{|J|=k}\prod_{j\in 
		J} x_j }. 
\end{align}
It is, of course, sufficient to prove the comparison for each
individual $k$. That is, it is enough to prove that for each
$k=1,\ldots, K$ we have
\begin{align} \label{e:comp.k}
\frac{\sum_{j=1}^K  x_j}{K}\geq& \frac{\frac{1}{{K
			\choose k}}   \sum_{|J|=k}\prod_{j\in 
		J} x_j}{\frac{1}{{K
			\choose k-1}}   \sum_{|J|=k-1}\prod_{j\in 
		J} x_j }. 
\end{align}

Note that for $k=1$, the inequality \eqref{e:comp.k} becomes an
identity. To prove it for $k=2,\ldots, K$, we denote
$$
S_k = \frac{1}{{K
		\choose k}}   \sum_{|J|=k}\prod_{j\in 
	J} x_j, \ k=2,\ldots, K,
$$
so that \eqref{e:comp.k}  states that
\begin{align} \label{e:comp.S}
S_1\geq \frac{S_k}{S_{k-1}}, \ k=2,\ldots, K.
\end{align}

The Maclaurin inequalities state that
\begin{equation} \label{e:macl}
S_1\geq S_2^{1/2}\geq \cdots \geq S_K^{1/K};
\end{equation}        
see e.g. \cite{cvetkovski2012newton}. It follows from \eqref{e:macl} that
for any $k=2,\ldots, K$ we have
\begin{align*}
&S_1S_{k-1}\geq S_{k-1}^{1/(k-1)}S_{k-1} = S_{k-1}^{k/(k-1)} \\
=& \bigl(
S_{k-1}^{1/(k-1)}\bigr)^k
\geq \bigl( S_k^{1/k}\bigr)^k = S_k,
\end{align*}
hence proving \eqref{e:comp.S}.
\end{proof}

\section{Proof of Result in Section \ref{sec:4}}

\begin{proof}[Proof of Theorem \ref{thm:general}]
    To prove the result, we rely on the following observations.

    \begin{itemize}
        \item[a] For an arbitrary node $i$ in the graph with degree at least 2, the difference between number of movements from it to any other different neighbour nodes will never exceed one.

        It then implies that, whenever a (non-last visited) node $i$ has been visited for at least $d_i k$ times, each of its neighbour nodes must have been visited for $k$ times. Here $d_i$ is the degree of node $i$.
        \item[b] Suppose node $i_0$ is the last visited node. Since the graph is connected, for any node $j \neq i_0$, there exists a path $(v_0,v_1,\ldots,v_{l_j})$ with $v_0 = i_0, v_{l_j} = j$ connects nodes $i_0$ and $j$.

        From observation a., we can deduce that whenever node $j$ has been visited for $T_j := 1\cdot d_{v_1} \cdot d_{v_2} \cdots d_{v_{l_j}}$, then node $i_0$ must be visited for at least once.
    \end{itemize}
    By above observations, we can compute that 
    \begin{eqnarray}\label{eqn:TC}
        T_C &\leq& 1 + \sum_{j \neq i_0} T_j \nonumber \\
        &\leq& 1 + \sum_{j \neq i_0} d_{max}^{l_j} \nonumber \\
        &\leq& 1 + (V - 1) \cdot d_{max}^{l_{max}}, 
    \end{eqnarray}
    where $d_{max} := \max_{j} d_j$, $V$ is the number of total nodes, $l_{max}$ is the length of longest path in the graph. Therefore, we conclude the proof by letting $G = 1 + (V - 1) \cdot d_{max}^{l_{max}}$.
\end{proof}

\begin{proof}[Proof of Theorem \ref{thm:star}]
	First of all, it can be seen that it must move to node 0 whenever it is in node $i \in \{1, \ldots, n\}$.
	For random walk policy, the question is then reduced to computing the expectation of numbers $N_0$ for visiting all child nodes 1 to $n$ starting from state 0.
	This becomes a standard Coupon collector's problem.
	It is known that $\mathbb E[N_0] = n (1 + \frac{1}{2} + \frac{1}{n})$. Then $\mathbb E_{\pi_{rw}}[T_C] = 2 \mathbb E[N_0] - 1
	= 2n (\sum_{i=1}^n \frac{1}{i}) - 1$, since all edges except the last one must be traversed in both directions .
	
	For negative-feedback policy, it can be observed that (i) node $i \in \{1,\ldots, n\}$ can be only visited if the previous state is 0; (ii) the node $i$ will be visited twice only if all states $j \neq i$ have been visited at least once. Therefore, we conclude that $\mathbb E_{\pi_{neg}}[T_C] = 2 n - 1$. 
\end{proof}

\begin{proof}[Proof of Theorem \ref{thm:path}]
	For any integer $n \geq 1$, we write $\mathbb E_{\pi}[T_C(n)]$ as the cover time of path with length $n$ under policy $\pi$ and we prove the statement by induction method.
	If $n = 2$, we can easily list all possible paths for negative-feedback policy.
	\begin{table}[ht!]
		\centering
		\begin{tabular}{ c|c|c } 
			\hline
			\hline
			& Path & Probability  \\
			\hline 
			1& (0,1,2) & 1/2 \\
			2& (0,1,0,1,2) &  1/2 \\
			\hline
		\end{tabular}
		\caption{In the path graph with $n=2$, all possible paths along with corresponding probability under negative feedback policy.}\label{tab:pathn2}
	\end{table}
   Then $\mathbb E_{\pi_{neg}}[T_C(2)] = 2 \times \frac{1}{2} + 4 \times \frac{1}{2} = 3 < 4$.
   Therefore, statement is true when $n = 2$.
   Suppose that $\mathbb E_{\pi_{neg}}[T_C(n)] < n^2$ for all $n \leq n_0$, we need to prove $\mathbb E_{\pi_{neg}}[T_C(n)] < n^2$ holds for $n = n_0 + 1$ as well.
   
   Note that the initial state is 0. After the first move, it will be in state 1.
   We then consider a restricted graph onto states $\{1,\ldots,n\}$ (i.e. remove state 0 and the edge connecting states 0 and 1).
   By induction, we know that $\mathbb E_{\pi_{neg}}[T_C'(n-1)] < (n-1)^2$,
   where $\mathbb E_{\pi_{neg}}[T_C'(n-1)]$ is the cover time in this restricted graph. 
   
   Next, we use the key observation in Lemma \ref{lem:key} that state 0 can be visited at most $n-1$ times before all states are visited.
   Therefore, it leads to $\mathbb E_{\pi_{neg}}[T_C(n)] \leq 1 + \mathbb E_{\pi_{neg}}[T_C'(n-1)] + 2(n-1) < 1 + (n-1)^2 + 2(n-1) = n^2$.
   Here, 1 corresponds to the first move, $2(n-1)$ corresponds to the fact that the edge between nodes 0 and 1 can be traversed in both direction for at most $n-1$ times.
   
   Finally, it is easy to get that $\mathbb E_{\pi_{rw}}[T_C(n)] = n^2$ for random walk policy.
   This concludes the proof.
\end{proof}

\begin{lemma}\label{lem:key}
	In path graph with $n \geq 1$, state 0 can be visited at most $n-1$ times (excluding time $0$) before it reaches state $n$.
\end{lemma}

\begin{proof}[Proof of Lemma \ref{lem:key}]
	For node $i \in \{1, \ldots, n-1\}$, we define $x_i(t) = N_{i,i-1}(t) - N_{i,i+1}(t)$, where $t$ is the index of number of time visiting state 0 and 
    $N_{i,j}(t)$ represents the number of moves from node $i$ to node $j$ ($j = i-1$ or $i+1$) up to $t$-th visit to initial state 0.
	When $t = 0$, $x_i(0) \equiv 0$ for any node $i$.
	Under negative feedback policy, $x_i(t)$ can only take value in $\{-1,0,1\}$.
	
	Moreover, considering any two consecutive visits to state 0. We examine changes in $x_i(t)$'s.  
	It is not hard to see that 
	$$\sum_{i \in \{1, \ldots, n-1\}} x_i(t+1) - \sum_{i \in \{1, \ldots, n-1\}} x_i(t) = 1.$$ 
	See Figure \ref{fig:pathproof} for more intuition.
	\begin{figure}
		\centering
		\mbox{
			\includegraphics[width=0.6\textwidth]{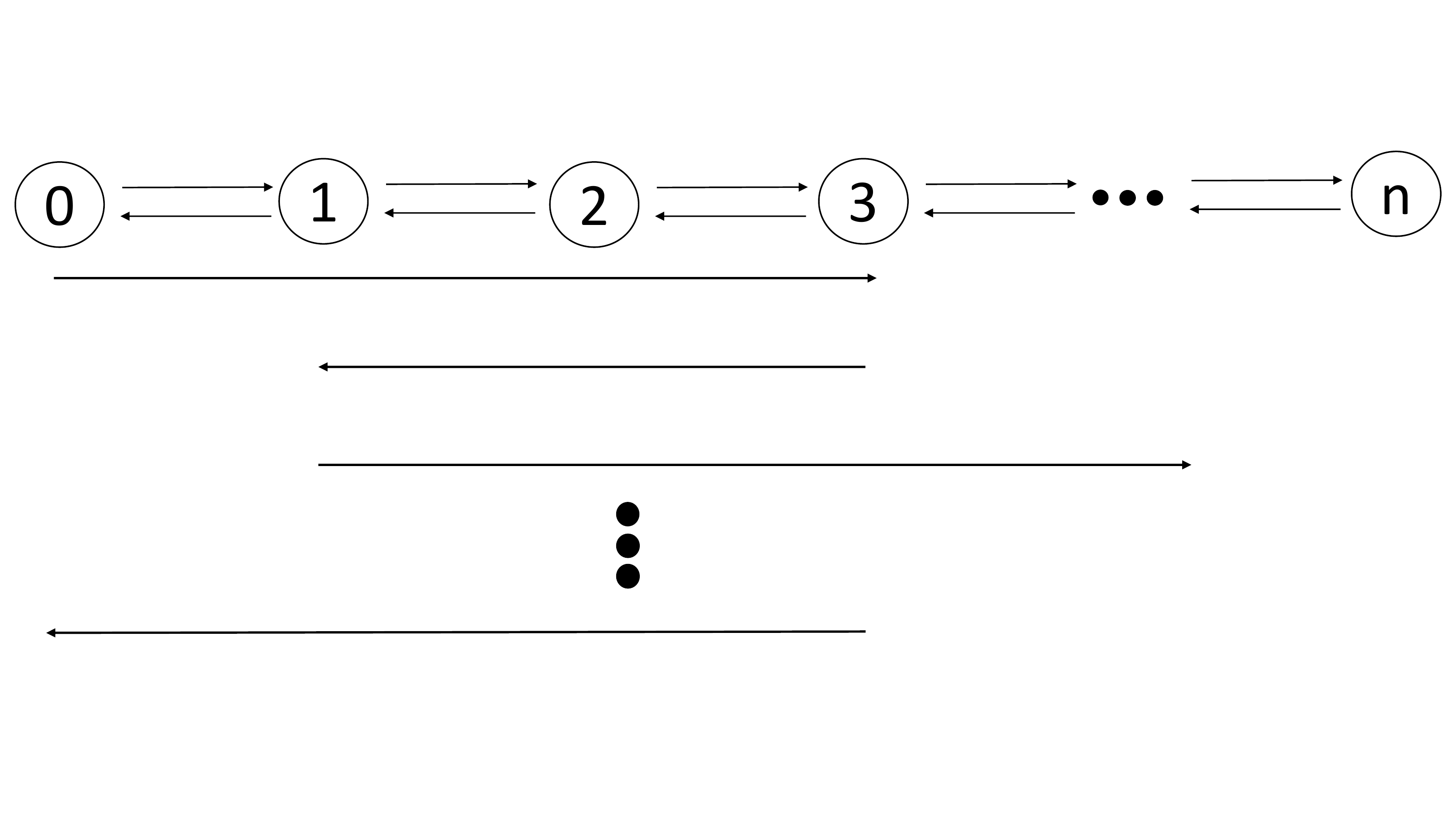}
		}	
		\caption{A graphical illustration of two consecutive visits to state 0 in the path graph. In above, the agent starts from state 0 and moves right until it reaches state 3. Then it moves left to get to state 1 and turn right again.
        The agent repeats taking left or right until it gets back to state 0.
        It helps understanding the relation $ \sum_{i \in \{1, \ldots, n-1\}} x_i(t+1) - \sum_{i \in \{1, \ldots, n-1\}} x_i(t) = 1$.}\label{fig:pathproof}
	\end{figure}
	
	Note that $\sum_{i \in \{1, \ldots, n-1\}} x_i(t)$ is upper bounded by $n-1$. At the start, $\sum_{i \in \{1, \ldots, n-1\}} x_i(0) = 0$.
	Therefore, $T_{max} \leq (n-1)/1 = n-1$, where $T_{max}$ is defined to be the number of times visiting state 0 before reaching state $n$.
	This completes the proof. 
\end{proof}

\begin{proof}[Proof of Theorem \ref{thm:circle}]
To prove the result, we need the following key observations.

\begin{itemize}
    \item[] \textbf{Obs} 1. Suppose node $k$ is the last visited node. Then the last move is either the edge from node $k-1$ to $k$ or from node $k+1$ to $k$. Without loss of generality, we can assume the last move is the edge from node $k-1$ to $k$; see Figure \ref{fig:circleproof} for graphical illustration.
    \item[] \textbf{Obs} 2. According to Lemma \ref{lem:key}, we know that the number of movements from node $0$ to node $1$ is no larger than $k$.
    \item[] \textbf{Obs} 3. It holds that the number of movements from node $0$ to node $n$ is no greater than $\min\{k, n-k\}$. 
    
    Reasons: (i) If the number is larger than $n-k$, then by Lemma \ref{lem:key}, the node $k$ must be visited from node $k+1$ to $k$, which contradicts with the definition of the last visited node.
    (ii) If the number is larger than $k$, which means that there are at least $k+1$ movements from node $0$ to $n$. However, the last movement at node 0 is the edge from node 0 to node 1 according to the definition of the last visited node and structure of circle. Furthermore, it is the $l$-th movement from node 0 to node 1 and $l \leq k$ (by \textbf{Obs} 2.)
    In other words, the agent takes edge from node 0 to node $n$ for $k+1$ times before it takes edge from node 0 to node 1. This conflicts with the mechanism of the negative feedback algorithm.
    
    \item[] \textbf{Obs} 4. The number of movements from node $n$ to $0$ is at most $\min\{k, n - k\}$.

    Reasons: it suffices to show that the number of movements from node $n$ to $0$ is always smaller or equal to the number of movements from node $0$ to $n$.
    This is true because that node $k$ is the last visited node and there is no movement between node $k+1$ and $k$. Therefore, it is impossible to happen that there is no movement from node 0 to node $n$ between two consecutive movements from node $n$ to 0.

    \item[] \textbf{Obs} 5. The number of movements from node $n'$ to $n' + 1$ ($k < n' < n$) is no greater than the number of movements from node $n' + 1$ to $n'$.

    Reason: the same logic as that of \textbf{Obs} 4.

    \item[] \textbf{Obs} 6. The number of movements from node $n'+1$ to $n'$ ($k < n' < n$) is no greater than $\min\{k + n - n', n'- k\}$.

    Reasons: the fact that the number is smaller than or equal to $n'-k$ again follows from Lemma \ref{lem:key}. 
    The number is smaller than $n' + 1$ can be proved by induction. 
    It suffices to show that the number of movements from node $n'+1$ to $n'$ is no greater than the number of movements from node $n'+2$ to $n'+1$ plus 1. 
    This is true because, by the mechanism of the negative feedback algorithm, it always holds that 
    the number of movements from node $n'+1$ to $n'$ is no greater than the number of movements from node $n'+1$ to $n'+2$ plus 1.
    The latter quantity is further smaller than or equal to the number of movements from node $n'+2$ to $n'+1$ plus 1 by \textbf{Obs} 5.
\end{itemize}

By the same logic, we further have the following observations.
\begin{itemize}
    \item[] \textbf{Obs} 7. The number of movements from node $n'$ to node $n' + 1$ ($0 \leq n' \leq k - 1$) is no larger than $\min\{k - n', n - k + 1 + n'\}$.
    \item[] \textbf{Obs} 8. The number of movements from node $n' + 1$ to node $n'$ ($0 \leq n' \leq k - 1$) is no greater than 
    $\min\{k - n' - 1, n - k + n'\}$.
\end{itemize}

Thanks to the above observations. 
Define $T_{right}$ to be the number of movements that the agent moves between node $0$ and node $k$ (i.e. the right part of circle) and 
$T_{left}$ to be the number of movements that the agent moves between node $0$ and node $k+1$ (i.e. the left part of circle).
We then know that 
\begin{eqnarray}\label{eq:circle}
    &&\mathbb E_{\pi_{neg}}[T_c] \nonumber \\
    &=& 
    \mathbb E_{\pi_{neg}}[T_{right}] + \mathbb E_{\pi_{neg}}[T_{left}] \nonumber \\
    &\leq& \mathbb E_{\pi_{neg}}[T_{right}] + \sum_k p_k \cdot 2 \sum_{k < n' \leq n} \min\{k + n - n', n'-k\} ~~(\text{using Obs. 5 and 6. })\nonumber \\
    &\leq& \sum_{k} p_k \cdot \nonumber ~~ (\text{where $p_k$ be the probability of node $k$ being the last visited node})\\
    &&\{ \underbrace{2 \sum_{0 \leq n' \leq k-1} \min\{k - n', n - k + 1 + n'\} - k + 2 \sum_{k < n' \leq n} \min\{k + n - n', n'-k\}}_{A_k} ~~ (\text{using Obs. 7 and 8.}) \} \nonumber \\
\end{eqnarray}
To compute the precise formula of \eqref{eq:circle}, we consider $n$ to be odd and even number separately.

When $n = 2n_1$ (even number), then 
$A_k$ in \eqref{eq:circle} becomes 
$2n_1^2 + k$ for $k \leq n_1$ and 
$2n_1^2 + 2n_1 - k$ for $k > n_1$.
Note that $p_k$ is non-zero for any $1 \leq k \leq n$, then \eqref{eq:circle} is strictly less than $2n_1^2 + n_1$.
On the other hand, $\mathbb E_{\pi_{rw}}[T_c] = 2 n_1^2 + n_1$. Hence, cover time of negative feedback algorithm is smaller than that of random walk policy for even $n$.

When $n = 2n_1 + 1$ (odd number), then 
$A_k$ in \eqref{eq:circle} becomes 
$2n_1^2 + 2n_1 + k$ for $k \leq n_1 + 1$ and 
$2n_1^2 + 4n_1 + 2 - k$ for $k > n_1 + 1$.
Hence, cover time of negative feedback algorithm is strictly smaller than $2n_1^2 + 3n_1 + 1$, which is exactly the cover time for random walk algorithm \cite{witter2022cover}. Therefore, we conclude the proof.

\begin{figure}
		\centering
		\mbox{	\includegraphics[width=0.8\textwidth]{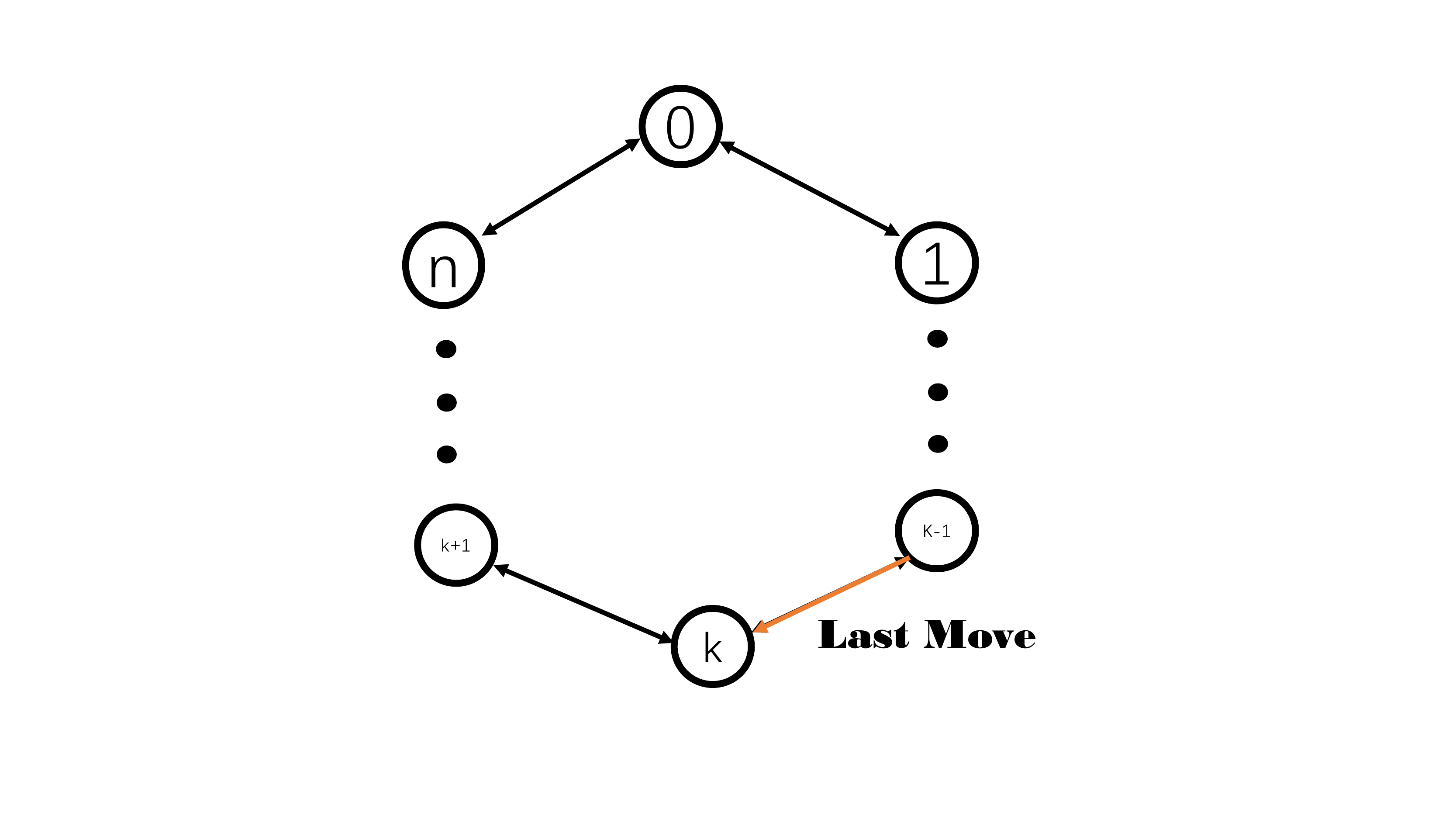}
		}	
		\caption{A graphical illustration of movements under the negative feedback policy with the last move from node $k-1$ to node $k$.}\label{fig:circleproof}
	\end{figure}
\end{proof}

\begin{proof}[Proof of Theorem \ref{thm:complete}]
	It is straightforward to see that 
	$\mathbb E_{\pi}[T_C] = 1 + \sum_{i=1}^{n-1} T_{i,i+1}^{\pi}$ for any policy $\pi$,
	where $T_{i,i+1}^{\pi}$ is the expectation of $N_{i,i+1}^{\pi}$, where the definition of latter quantity is the first time to reach the $(i+1)$-th un-visited node starting from $i$-th un-visited node.
	
	In random walk policy, $N_{i,i+1}^{\pi_{rw}}$ is a random variable following geometric distribution with parameter $\frac{n-i}{n-1}$. Then $T_{i,i+1}^{\pi_{rw}} = \frac{n-1}{n-i}$.
	
	In negative-feedback policy, during each move from a visited node $j$, the success probability of visiting an un-visited node is at least $ \frac{n-i}{n-1}$. This is because any edge between node $j$ and un-visited nodes has not been traversed yet. 
	Therefore $(n-i)$ un-visited nodes must belong to set $Smin_j^{(n)}$.
	In addition, cardinality $Kmin_j^{(n)}$ is trivially upper bounded by $n-1$. Therefore, success probability is at least $(n-i)/(n-1)$. 
    Thus $N_{i,i+1}^{\pi_{neg}}$ is stochastically smaller than $N_{i,i+1}^{\pi_{rw}}$ which gives
    $ T_{i,i+1}^{\pi_{neg}} = \mathbb E[N_{i,i+1}^{\pi_{neg}}] \leq \frac{n-1}{n-i}$.
    (Here we say a random variable $X$ is stochastically smaller than random variable $Y$, if their cumulative distribution functions satisfy $F_X(t) \geq F_Y(t)$ for any $t \in \mathbb R$.)
	
	Next, we need to show that there is at least one $i$ such that 
	$T_{i,i+1}^{\pi_{neg}} < \frac{n-1}{n-i}$ holds. We prove this by contradiction method.
	If not, then $T_{i,i+1} = \frac{n-1}{n-i}$ for all $i \in \{1, \ldots, n-1\}$.
	This implies that each node $j$ has been visited at most once.
	This is due to the fact that the success probability will be at least $(n-i_{j,2})/(n-2)$, when $j$ is visited for the second time.
	Here $i_{j,2}$ represents the number of visited nodes before $j$ is visited twice. Then $T_{i_{j,2},i_{j,2}+1}^{\pi_{neg}} < \frac{n-1}{n - i_{j,2}}$ which violates our assumption that $T_{i,i+1} = \frac{n-1}{n-i}$ for all $i$.
	Hence, all states must be visited for at most one time.
	On the other hand, $\mathbb E_{\pi_{neg}}[T_C]
	= \sum_{i} \frac{n-1}{n-i} > n$ which indicates that, with non-zero probability, at least one node has been visited for multiple times.
	This leads to the contradiction.
	
	Therefore, $\mathbb E_{\pi_{neg}}[T_C] < \mathbb E_{\pi_{rw}}[T_C]$ holds. 
\end{proof}

\begin{proof}[Proof of Theorem \ref{thm:tree}]
    We first show that the root node is visited at most $bH$ times before all nodes have been visited at least once.
    To prove this, we rely on the following observation (Lemma \ref{lem:obs}).
    \begin{lemma}\label{lem:obs} 
    The root node is visited at most $bh$ times before all nodes at depth $h$ have been visited at least once.
    \end{lemma}
    \begin{proof}[Proof of Lemma \ref{lem:obs}]
        We take an arbitrary node $i_h$ at depth $h$. Thanks to the tree structure, we know that there exists a unique path that connects the root node $i_0$ and intermediate node $i_h$ and the length of this path is exactly $h$.
        For notational simplicity, we write this path as 
        $\{i_0, i_1, i_2, \ldots, i_h\}$. 
        Similar to the proof of Lemma \ref{lem:key}, we can define $x_j(t)
        = N_{i_j,i_{j-1}}(t) - N_{i_j,i_{j+1}}$(t)
        where $t$ is the index of number of time visiting state 0 and 
        $N_{i_j,i_{j-1}}(t)$ ($N_{i_j,i_{j+1}}(t)$) is the number of moves from node $i_j$ to node $i_{j-1}$ ($i_{j+1}$). 
        By repeating the rest of proof of Lemma \ref{lem:key}, we arrive at the conclusion there is at most $h$ moves from $i_1$ to $i_0$ before node $i_h$ is reached. 
        Since root node $i_0$ has at most $b$ children. Therefore, root node could be visited at most 
        $b h$ times before all nodes at depth $h$ have been visited once. This concludes this lemma.
    \end{proof}
    We especially take $h = H$ in Lemma \ref{lem:obs} and conclude that the root node is visited at most $bH$ times.

    Next, we prove that the node at depth $h$ could be visited at most $(b+1) (H + h)$ times before all nodes are visited at least once. 
    By previous argument, we have proved that there are at most $H$ moves from root node $i_0$ to node $i_i$ ($i_1$ is an arbitrary node at depth 1)
    and $H$ moves from node $i_1$ to root node $i_0$.
    By the property of negative feedback algorithm, there are at most $H + 1$ moves from node $i_1$ to each of its children $i_2$.
    Hence, the number of moves from its arbitrary child $i_2$ to node $i_1$ is at most $H+1$.
 (This is due to the another observation that the number of moves from child node to parent node is no larger than the number of moves from parent node to child node.) 
 To sum up, node $i_1$ at depth 1 can be visited at most $(b+1)(H+1)$ times.
 By repeating such procedure, we can have that the number of moves from node $i_{h+1}$ at depth $h+1$ to its parent node $i_h$ is at most $H + h$
 and the number of moves from node $i_{h-1}$ to its child node $i_h$ is at most $H + h - 1$.
 Therefore, node $i_h$ at depth $h$ can be visited at most $(b+1) (H + h)$ times.
 By taking $h = H$, we finally conclude that each node is visited for at most $2(b+1)H$ times before it covers all nodes.
\end{proof}

\begin{proof}[Proof of Theorem \ref{thm:tree:all}]
    In the proof of Theorem \ref{thm:tree}, we have already shown that the number of moves from node $i_{h-1}$ (at depth $h-1$) to its child node is at most $H + h - 1$.
    For any leaf node, we know it is at depth $H$.
    Therefore, there is at most $H + H - 1 \leq 2H$ number of times to visit a single leaf node.

    Therefore, the cover time is at most 
    \begin{eqnarray}
        T_C &\leq& \#\{\text{non-leaf nodes}\} 2(b+1)H + \# \{\text{leaf nodes}\} 2H \nonumber\\
        &\leq& 2(b+1)H \cdot \frac{b^{H} - 1}{b-1}
        +2 H b^H \nonumber \\
        &\leq& 4 H \frac{b+1}{b-1} b^H.
    \end{eqnarray}
    This completes the proof.
\end{proof}

\end{document}